\documentclass[11pt,letterpaper]{article}

\usepackage[includeheadfoot,
            marginratio={1:1,2:3}, 
            width=412pt,
            height=688pt,]{geometry}
            \usepackage{tikz}
\usepackage[compat=1.1.0]{tikz-feynman}
\usepackage[utf8]{inputenc}
\usepackage{amsmath}
\usepackage{amsfonts}
\usepackage{amssymb}
\usepackage{graphicx}
\usepackage{booktabs}
\usepackage{empheq}
\usepackage{paralist}
\usepackage{cite}
\usepackage[hidelinks]{hyperref}
\usepackage{xcolor}
\usepackage{tensor}
\usepackage{adjustbox}
\usepackage{subcaption}
\usepackage{braket}
\usepackage{tikz}
\usepackage{tikz-cd}
\usepackage{yfonts}
\usepackage{enumitem}
\usepackage{physics}
\usepackage{mdframed}

\usetikzlibrary{decorations.pathreplacing,calc}
\newcommand{\nc}{\newcommand}
\nc{\lb}{\llbracket}
\nc{\rb}{\rrbracket}
\nc{\gl}{\llbracket}
\nc{\gr}{\rrbracket}
\nc{\bbR}{\mathbb{R}}
\nc{\bbC}{\mathbb{C}}
\nc{\bbZ}{\mathbb{Z}}
\nc{\cO}{\mathcal{O}}
\nc{\cS}{\mathcal{S}}
\nc{\cM}{\mathcal{M}}
\nc{\cT}{\mathcal{T}}
\nc{\cX}{\mathcal{X}}
\nc{\cQ}{\mathcal{Q}}
\nc{\cD}{\mathcal{D}}
\nc{\cC}{\mathcal{C}}
\nc{\cG}{\mathcal{G}}
\nc{\cF}{\mathcal{F}}
\nc{\cI}{\mathcal{I}}
\nc{\pd}{\partial}
\nc{\la}{\lambda}
\newcommand\beq{\begin{equation}}
\newcommand\eeq{\end{equation}}
\nc{\del}{\partial}
\nc{\tri}{\hspace{-3.5pt}\vartriangle\hspace{-3.5pt}}
\nc{\blacktri}{\blacktriangle}
\nc{\eq}[1]{\begin{equation}
                     \begin{split} #1 \end{split}
                     \end{equation}}
\nc{\ul}{\underline}
\nc{\ov}{\overline}
\nc{\fa}{\hat}
\nc{\fb}{\MakeUppercase}
\nc{\fc}{\tilde }
\nc{\Lie}{{\cal L}} 
\nc{\lambdabar}{{\mkern0.75mu\mathchar '26\mkern -9.75mu\lambda}}

\newmuskip\pFqmuskip

\newcommand*\pFq[7][8]{
  \begingroup % only local assignments
  \pFqmuskip=#1mu\relax
  \mathchardef\normalcomma=\mathcode`,
  \mathcode`\,=\string"8000
  \begingroup\lccode`\~=`\,
  \lowercase{\endgroup\let~}\pFqcomma
  {}_{#2}{#3}_{#4}{\left[\left.\genfrac..{0pt}{}{#5}{#6}\right|#7\right]}
  \endgroup
}
\newcommand{\pFqcomma}{{\normalcomma}\mskip\pFqmuskip}

\allowdisplaybreaks[2]

\begin{document}

\vspace*{1.5cm}
\begin{center}
{\huge
Complexity in Tame Quantum Theories}
%\\[.4cm]
%{\Large $\sharp$o-minimality}

%\vspace{.6cm}
\end{center}

\vspace{0.35cm}
\begin{center}
 Thomas W.~Grimm,
 Lorenz Schlechter, 
 Mick van Vliet
\end{center}

\vspace{.5cm}
\begin{center} 
\vspace{0.25cm} 
\emph{
Institute for Theoretical Physics, Utrecht University,
\\
Princetonplein 5, 3584 CC Utrecht, 
The Netherlands } \\
\end{center}

\vspace{1.5cm}

%%%%%%%%%%%%%%%%%%%%%%%%%%%%%%%%%%%%%%%%%%%%%%%
%%%%%%%%%%%%%%%%%%%%%%%%%%%%%%%%%%%%%%%%%%%%%%%
%%%%%%%%%%%%%%%%%%%%%%%%%%%%%%%%%%%%%%%%%%%%%%%
%%%%%%%%%%%%%%%%%%%%%%%%%%%%%%%%%%%%%%%%%%%%%%%
%%%%%%%%%%%%%%%%%%%%%%%%%%%%%%%%%%%%%%%%%%%%%%%
%%%%%%%%%%%%%%%%%%%%%%%%%%%%%%%%%%%%%%%%%%%%%%%
%%%%%%%%%%%%%%%%%%%%%%%%%%%%%%%%%%%%%%%%%%%%%%%
%%%%%%%%%%%%%%%%%%%%%%%%%%%%%%%%%%%%%%%%%%%%%%%

\begin{abstract}
\noindent
Inspired by the notion that physical systems can contain only a finite amount of information or complexity, 
we introduce a framework that allows for quantifying the amount of logical information needed to specify a function or set. 
We then apply this methodology to 
a variety of physical systems and derive 
the complexity of parameter-dependent physical observables and coupling functions appearing in effective Lagrangians. 
In order to implement these ideas, it is essential to consider physical theories that can be defined in an o-minimal structure. O-minimality, a concept from mathematical logic, encapsulates a tameness principle. 
It was recently argued that this property is inherent to many known quantum field theories and is linked to the UV completion of the theory. 
To assign a complexity to each statement in these theories one has to further constrain the allowed o-minimal structures. 
To exemplify this, we show 
that many physical systems can be formulated using Pfaffian o-minimal structures, which have a well-established notion of complexity.  
More generally, we propose adopting sharply o-minimal structures, recently introduced by Binyamini and Novikov, as an overarching framework to measure complexity in quantum theories. 
\end{abstract}

\clearpage

\tableofcontents

%%%%%%%%%%%%%%%%%%%%%%%%%%%%%%%%%%%%%%%%%%%%%%%
%%%%%%%%%%%%%%%%%%%%%%%%%%%%%%%%%%%%%%%%%%%%%%%
%%%%%%%%%%%%%%%%%%%%%%%%%%%%%%%%%%%%%%%%%%%%%%%
%%%%%%%%%%%%%%%%%%%%%%%%%%%%%%%%%%%%%%%%%%%%%%%
%%%%%%%%%%%%%%%%%%%%%%%%%%%%%%%%%%%%%%%%%%%%%%%
%%%%%%%%%%%%%%%%%%%%%%%%%%%%%%%%%%%%%%%%%%%%%%%
%%%%%%%%%%%%%%%%%%%%%%%%%%%%%%%%%%%%%%%%%%%%%%%
%%%%%%%%%%%%%%%%%%%%%%%%%%%%%%%%%%%%%%%%%%%%%%%

\newpage

\parskip=.2cm
\section{Introduction}
Over the years, physicists have pondered the idea that physical systems might have a fundamental limit to their information content or complexity.
In simple scenarios, realizing this notion appears straightforward -- consider, for instance, a system characterized by a finite number of qubits. However, the issue becomes much more involved if one considers more intricate systems. Central to this challenge is the question of how one might reconcile such a principle with the formulation of quantum mechanics, which often posits infinite-dimensional Hilbert spaces, or the vast expanse of higher-dimensional quantum field theories. This paper introduces a novel approach that defines the notion of finite complexity within a conceptually clean and general mathematical formalism. Our focus lies on understanding the geometric or functional complexity of physical quantities, and the amount of information needed to describe these quantities in terms of physical parameters. In doing so, we connect with recent mathematical advancements that define complexity for sets and functions within certain `tame' logical structures.

Tameness, a concept from mathematical logic, introduces a generalized notion of finiteness \cite{VdDries}. When applied to sets in real Euclidean space, tameness singles out those sets with finitely many connected components. This seemingly simple requirement gains depth when also mandating that the collection of tame sets is closed under any finite combination of the usual set-theoretic operations as well as linear projections. Consequently, one can define tame functions, having graphs which are tame sets, to pave the way for a new notion: a tame geometry. The described principle of tameness is also known as o-minimality and has led to a framework rich with powerful theorems and analytical tools. 

Only recently it was observed \cite{Grimm:2021vpn,Douglas:2022ynw,Douglas:2023fcg} that tame geometry arises in various physical settings, ranging from observables in quantum field theories to the landscape of effective field theories consistent with quantum gravity. 
The primary objective of this work is twofold: firstly, to deepen the connection between tame geometry and physical systems by analyzing simple and extendable examples of tame quantum systems; and secondly, to present a sharpened notion 
of o-minimality. This stronger version of o-minimality captures the idea that sets and functions should admit a well-defined, finite complexity. Furthermore, the presence of a natural measure of complexity promotes tameness from a qualitative to a quantitative feature. Remarkably, it turns out to be present in all the physical examples in which o-minimality has been analyzed so far. This leads us to suggest that this notion of `sharp o-minimality' introduced by Binyamini and Novikov \cite{BinNovICM,binyamini2022sharply} is the tameness principle inherent to physics.

Connections between tame geometry and the ideas of complexity have a long history. For example, Khovanskii's theory of fewnomials \cite{Khovanskii} notably influenced foundational works on o-minimal structures, such as \cite{WilkieRExp}. The focus of this paper's initial parts heavily leans on Gabrielov and Vorobjov's works \cite{GabVor95,GabVor01,GabVor04}, which define complexity for Pfaffian functions. 
These functions are polynomials of some variables and a finite set of analytic functions, termed the Pfaffian chain, which satisfy a system of first-order differential equations that is again only defined by polynomials. 
Due to these polynomial-based equations, Pfaffian functions possess finiteness properties comparable to polynomials despite the fact that one can use transcendental functions, such as the exponential function, as part of the Pfaffian chain.  
We will describe in detail how to assign a complexity to Pfaffian functions and how it determines a computational and topological complexity of the sets generated by these functions.

We will argue that Pfaffian functions have numerous appearances in physics, which allows us to test this notion of complexity in various physical settings. Firstly, we consider lattice quantum field theory, where we explicitly compute the complexity of the correlation functions using contemporary techniques for Feynman integrals \cite{Weinzierl:2022mmp}. Secondly, we look at simple quantum mechanical systems. In particular, we evaluate the complexity for states of the harmonic oscillator and comment on the relation to other notions of complexity for quantum states \cite{Chapman:2021jbh}. Finally, we compute the complexity of gauge coupling function in Seiberg-Witten theory \cite{Seiberg:1994rs} and discuss how the complexity depends on the rank of the gauge group. 

While the theory of Pfaffian functions provides a straightforwardly computable measure of complexity that is applicable to several examples, it has a number of shortcomings.
From a physics perspective, we find that it is too restrictive when considering a broader choice of physical systems and observables. From a mathematical point of view it also does not fully elucidate the connection between tameness and complexity. The big leap in generalizing the Pfaffian construction was recently made by a refinement of the o-minimality axioms and the introduction of sharp o-minimality \cite{BinNovICM,binyamini2022sharply}, or $\sharp$o-minimality. Unlike the theory of Pfaffian functions, sharp o-minimality allows one to compute the complexity of every tame object and thereby provides a significantly more general framework of tameness and complexity. It naturally leads us to introduce the notion of sharp complexity, or $\sharp$complexity, which assigns to each set or function a finite list of tuples $(F,D)$, measuring their information content in different representations. This novel notion is compatible with all logical operations, but poses some interesting interpretational challenges that we will discuss. 

Due to the fact that sharp o-minimality was only recently introduced, there are only few established examples of such structures. Therefore, we will need to rely on the conjectures of \cite{BinNovICM} for concrete physical applications. Assuming these 
conjectures we study the sharp-ominimality of a broader class of quantum field theories on a finite lattice and quantum mechanical systems with general polynomial potentials. Asserting that period integrals 
are definable in a sharp o-minimal structure \cite{BinNovICM}, we will be able to discuss  general SU$(N)$ Seiberg-Witten theories. Moreover, we will conclude that the finite-loop amplitudes are not only o-minimal \cite{Douglas:2022ynw}, but in fact sharply o-minimal. These observations imply that physical quantities in all these settings should admit a well-defined complexity and can be analyzes using the recent strong theorems about sharp o-minimal structures. 

The outline of this work is as follows. In section~\ref{sec:tameness} we give a very brief introduction to tameness and o-minimal structures, also commenting on some of the main examples of structures relevant to this work. In section~\ref{Pfaff_definability} we give a more thorough introduction to Pfaffian structures and the theory of Pfaffian functions. In this context, we also introduce a first notion of complexity and discuss several of its important properties. We then implement this into physics in section~\ref{sec:quantumsystems}, where we compute and analyze the complexity in three physical settings: quantum field theories on points, quantum mechanical systems, and Seiberg-Witten theory. We will observe in each of these examples, that the Pfaffian structure is too restrictive when aiming at a complete description of all theories of these types. This leads us to introduce 
sharply o-minimal structures in section~\ref{sec:sharp-ominimal}. We will motivate this generalization, introduce the accompanying notion of sharp complexity, and summarize concrete candidate examples of such structures following \cite{BinNovICM}. In the final section~\ref{sec:sharp_physics} we revisit the physical settings and argue that sharp o-minimality now provides a sufficiently rich setting.

\section{Tameness and o-minimal structures}\label{sec:tameness}

In this section we give a very brief introduction to tame geometry and o-minimal structures. We will review some elementary concepts and list the examples of o-minimal structures that are most relevant to this work. The reader who is not familiar with the subject may additionally consult the basic introduction of van den Dries~\cite{VdDries}.  

\subsection{Tame sets and functions}
\label{sec:basics_o-min_1}

The starting point for tame geometry is to define what it means for a subset of Euclidean space to be tame. For consistency, the collection of tame sets should be closed under reasonable operations which one can perform on sets, such as taking unions and intersections. In mathematical terms, this collection should form a \textit{structure}. We denote a structure by $\cS=(\cS_n)$, where $\cS_n$ is a collection of subsets of $\bbR^n$. The axioms for a structure are:
\vspace*{-0.2cm}
\begin{itemize}
    \item[(i)] $\mathcal{S}_n$ is closed under finite intersections, finite unions, and complements\footnote{Here the complement is taken with respect to $\bbR^n$.};
     \item[(ii)] $\cS$ is closed under Cartesian products: $A \times B \in \cS_{n+m}$ if $A \in \cS_n$, $B \in \cS_m$; 
     \item[(iii)] $\cS$ is closed under any linear projection $\pi:\bbR^{n+1} \rightarrow \bbR^n$: $\pi(A)\in \cS_n$ if $A \in \cS_{n+1}$;
     \item[(iv)] $\cS_n$ contains the zero sets of all polynomials in $n$ real variables.
\end{itemize} 
\vspace*{-0.2cm}
The sets inside a structure are called \textit{definable}, to reflect the idea that they can be defined by means of simple set-theoretic operations. To ensure that definable sets are tame, an additional axiom is needed. This axiom defines an \textit{o-minimal structure}:
\vspace*{-0.2cm}
\begin{itemize}
     \item[(v)] the definable subsets of $\bbR$ have a finite number of connected components.\footnote{In other words, these sets are finite unions of intervals and points.}
\end{itemize}
\vspace*{-0.2cm}
This simple condition has strong consequences for the geometry of sets inside an o-minimal structure, and it conspires with axioms (i)-(iv) to create a remarkably constrained geometric framework. Even though the o-minimality axiom is only imposed on subsets of $\bbR$, the condition that structures are closed under linear projections ensures that higher-dimensional definable sets are constrained as well.

Within an o-minimal structure, there is a natural definition for tame functions. Given two definable sets $A$ and $B$ and a function $f:A\to B$, we will say that $f$ is definable if its graph $\Gamma(f)$ is a definable subset of $A\times B$. Because the graph of such a function is a tame set, definable functions exhibit similar tameness features to those of definable sets.  

The tameness of o-minimal structures can be understood through the lens of various theorems for definable sets and functions. The most essential of these is the cell decomposition theorem, which due to its technical nature we only state informally \cite{VdDries}. One first defines a simple type of set, called a cell, which can intuitively be thought of as a cube whose sides are deformed by means of tame functions. Importantly, one demands that linear projections of cells are also cells. The cell decomposition theorem now states that any definable set in an o-minimal structure can be decomposed into finitely many cells. In this sense, cells provide the elementary building blocks of tame sets. The power of the cell decomposition comes from the fact that the geometry of cells is extremely simple to understand, so that any geometric question concerning a tame set can always be reduced to finitely many simpler question about cells.

\subsection{Examples of o-minimal structures}
\label{sec:basics_o-min_1}

So far we have only discussed o-minimal structures abstractly, so let us provide a few important examples. Recall from axiom (iv) that every structure contains at least the algebraic sets, i.e. the zero sets of polynomials. The smallest structure consists of all the sets which can be obtained by applying the operations in axioms (i)-(iii) finitely many times, and it is denoted by $\bbR_\text{alg}$. 

The most natural way to obtain larger structures is as follows. One starts with a collection of functions $\cF$ which one would like to be definable, and then considers the structure \textit{generated by} $\cF$, denoted by $\bbR_\cF$. This means that $\bbR_\cF$ should contain the graphs of all the functions in $\cF$, and in general the sets in this structure can be thought of as all possible sets which can be obtained by applying the basic set-theoretic operations to these graphs finitely many times. Depending on the tameness of the functions in $\cF$, the resulting structure $\bbR_\cF$ may be o-minimal or not. 

The collection of functions $\cF$ may take many forms. For instance, one can consider to add a single function, such as the real exponential $\text{exp}:\bbR\to \bbR$. This leads to one of the most central o-minimal structures, denoted by $\bbR_\text{exp}$ \cite{WilkieRExp}. Intuitively, the sets in $\bbR_\text{exp}$ can be though of as being generated by zero sets of exponential polynomials, i.e.~equations of the form
\begin{equation}
    P(x_1,\ldots,x_n,e^{x_1},\ldots,e^{x_{n}})=0
\end{equation}
for some polynomial $P$. Another natural class of functions are the restricted analytic functions, which are defined as analytic functions restricted to compact domains.\footnote{This restriction is crucial for o-minimality, since analytic functions defined on unbounded domains are often not tame (e.g.~the sin function).} The structure generated by the exponential function together with the restricted analytic functions, denoted by $\bbR_{\text{an,exp}}$, was proven to be o-minimal in \cite{VdDriesMiller}. It plays a substantial role in tame geometry, most notably because period functions are definable in this structure \cite{BKT}. At present there are many known o-minimal structures and some of these will play a significant role in this work. Deferring a precise 
definition to later parts of the paper, let us list some further examples:
\begin{itemize}
 \item $\bbR_{\rm Pfaff}$. This structure is generated by solutions to certain differential equations, called Pfaffian functions. This structure is central to this work and will be discussed in detail in the next section.
 \item $\bbR_{\rm rNoether}$. Upon slightly weakening the conditions defining Pfaffian functions, one obtains the class of Noetherian functions. When restricting the domains to bounded sets, these functions are tame and generate an o-minimal structure. We will revisit these functions in section~\ref{sec:sharp-ominimal}.
 \item $\bbR_{\rm Qf}$. This structure is generated by Q-functions, which arise as solutions to geometric differential equations with sufficiently regular singularities. These frequently appear in geometry and physics, and will be used in section~\ref{sec:sharp-ominimal}.
\end{itemize}
One key point of this work is that the listed structures are not only o-minimal, but admit \textit{additional} properties that make the definition of a notion of complexity possible; they are expected to be examples of sharply o-minimal structures, to be defined in section~\ref{sec:sharp-ominimal}.\footnote{We will discuss the sharp o-minimality of these structures in section~\ref{sharpConjectures} referring to recent mathematical conjectures.} 
We will find that all considered physical quantities are describable within these more restricted class of structures. Note that, as suggested in  \cite{Douglas:2023fcg}, one can also turn the story around and choose as $\cF$ a set of functions arising in a physical system. For example, one 
can inquire about the tameness of the structures generated by the set of correlation functions in a given collection of quantum field theories.

\section{Pfaffian structures and Pfaffian complexity}
\label{Pfaff_definability}
The purpose of this section is to introduce a particular class of o-minimal structures -- the Pfaffian structures -- and explain how they allow for defining a precise notion of complexity. After we have defined these structures, we illustrate how the resulting Pfaffian complexity relates to other notions of complexity, such as topological and computational complexity. 

\subsection{Tame structures from Pfaffian chains}
\label{sec:pfaffian_chains}

\subsubsection*{Pfaffian chains}
As we have seen in the previous subsection, the interesting examples of o-minimal structures are obtained by extending $\bbR_\text{alg}$ by a set of functions of interest. In the following we introduce a special class of o-minimal structures that are obtained by including solutions to certain differential equations, known as \textit{Pfaffian chains}. To set up the definition, let $U\subseteq \bbR^n$ be an open box, i.e. a product of open intervals. A finite set of functions $\zeta_1,\ldots,\zeta_r:U\to\bbR$ is said to form a Pfaffian chain if it satisfies a system of first-order differential equations of the form 
\begin{equation}
\frac{\partial \zeta_i}{\partial x_j} =  P_{ij} (x_1,\ldots,x_n,\zeta_1,\ldots,\zeta_i)  \quad\text{for} \quad i=1,\ldots,r, \quad j=1,\ldots, n,
\end{equation}
where $P_{ij}$ is a polynomial in $n+i$ variables. Crucially, this system has a certain triangularity requirement, namely that the differential equations for $\zeta_i$ only involve the functions $\zeta_1,\ldots,\zeta_i$, and exclude $\zeta_{i+1},\ldots,\zeta_r$. If the box $U$ is bounded, i.e.~none of the intervals extends to infinity, then the Pfaffian chain is \textit{restricted}. The number of functions $r$ is called the \textit{order} of the Pfaffian chain, and its \textit{degree} $\alpha$ is the highest degree among the polynomials $P_{ij}$. We will use the symbol $\zeta=(\zeta_1,\ldots,\zeta_r)$ to denote the entire Pfaffian chain. Given such a chain formed by functions $\zeta_1,\ldots,\zeta_r$, a \textit{Pfaffian function} is a function $f:U\to\bbR$ of the form
\begin{equation}\label{eq:Pfaffianfunction}
f(x_1,\ldots,x_n) = P(x_1,\ldots,x_n,\zeta_1,\ldots,\zeta_r),
\end{equation}
where $P$ is a polynomial in $n+r$ variables. 
\subsubsection*{Pfaffian complexity}
There are several essential characteristic numbers involved in the definition of a Pfaffian function. These are 
\begin{itemize}
\item the number of variables $n$;
\item the order $r$ of the underlying Pfaffian chain $r$;
\item the degree $\alpha$ of the underlying Pfaffian chain;
\item the degree $\beta$ of the polynomial $P$ used to define $f$.
\end{itemize}
Together, these numbers define the \textit{Pfaffian complexity} of a Pfaffian function, and we write
\begin{equation}
\cC_\zeta(f) = (n,r,\alpha,\beta).
\end{equation}
In case that no confusion can arise, we will often simply refer to this quantity as the complexity of $f$. The subscript $\zeta$ emphasizes the dependence of the complexity on the underlying Pfaffian chain $\zeta$. Since there is no unique chain which describes a given Pfaffian function, we often omit the explicit dependence on the chain. Note that, unlike certain other measures of complexity, this quantity is not a single number. Therefore, Pfaffian complexity does not admit a natural order, so that the notion of optimal complexity is subtle. We will comment on the interpretation of this feature briefly after the examples and in detail in section \ref{sec:sharp-ominimal}. 

Below we list some examples of Pfaffian functions and their Pfaffian complexities.
\begin{enumerate}[label=(\roman*)]
\item An $n$-variable polynomial $Q$ of degree $\beta$ is a Pfaffian function of complexity $(n,0,1,\beta)$, as it can be defined using an empty Pfaffian chain.\footnote{Note that, even though there is no Pfaffian chain in this example, the degree $\alpha$ is defined to be greater than or equal to $1$ by convention.}
\item The exponential function $f(x)=e^{ax}$ is a Pfaffian function of complexity $(1,1,1,1)$, since it fits into the Pfaffian chain 
\begin{equation}
    \frac{\partial \zeta}{\partial x} = a \zeta \,.
\end{equation}
\item The function $f(x)=1/x$, defined on the interval $(0,\infty)$, has complexity $(1,1,2,1)$, since it is a solution to the equation
\begin{equation}
    \frac{\partial \zeta}{\partial x} = - \zeta^2.
\end{equation}
\item The function $f(x)=\cos(x)$ defined on the interval $(-\pi,\pi)$ can be obtained as a Pfaffian function as follows. Consider the chain $\zeta$ formed by
\begin{align}
\frac{\partial \zeta_1}{\partial x} &=  \frac{1}{2}\big(1+ \zeta_1^2 \big), \\
\frac{\partial \zeta_2}{\partial x} & = - \zeta_1\zeta_2, \nonumber
\end{align}
which has solutions $\zeta_1(x) = \tan(x/2)$ and $ \zeta_2(x) = \cos^2(x/2)$. Using trigonometric relations, one may now obtain $f$ as the Pfaffian function
\begin{equation}
f(x) = \cos(x) = 2\zeta_2(x) -1, 
\end{equation}
which shows that $\cC_\zeta(f) = (1,2,2,1)$. For any integer $m\geq 1$, the function $f_m(x)=\cos(mx)$ can be written as a polynomial of degree in $m$ in $f(x)=\cos(x)$, and therefore it has complexity $\cC_\zeta(f_m)=(1,2,2,m)$. These functions frequently appear in physics, e.g.~as vibrational modes of a string and as Kaluza-Klein modes in a circle compactification. 
\end{enumerate}

As already indicated by the notation, the complexity of a Pfaffian function depends on the underlying chain. To illustrate this, consider the mononomial $f(x)=x^d$. We consider two ways of constructing $f$ as a Pfaffian function. As in example (i) above, the most obvious construction is to let the Pfaffian chain be empty, denoted by $\zeta= \emptyset$, in which case we have $\cC_\emptyset(f) = (1,0,1,d)$. Alternatively, we can construct $f$ by means of differential equations, and consider the following Pfaffian chain $\zeta$ on the domain $U=(0,\infty)$,
\begin{align}
\frac{\partial \zeta_1}{\partial x} &=  - \zeta_1^2, \\
\frac{\partial \zeta_2}{\partial x} & =  d\, \zeta_1\zeta_2. \nonumber 
\end{align}
The solutions to this Pfaffian chain are $\zeta_1(x) = 1/x$ and $\zeta_2(x)=f(x)$, and the complexity of $f$ is now given by $\cC_\zeta(f) = (1,2,2,1)$. More generally, if $f$ is a degree $d$ polynomial consisting of $M$ monomial terms, we can add each monomial to the set of differential equations to obtain a Pfaffian chain $\zeta$ of length $r=M+1$. The complexity of the polynomial $f$ is then
\begin{equation}
    \cC_\zeta(f) = (n,M+1,2,1)\, .
\end{equation}
If $f$ is a \textit{fewnomial}, i.e.~a polynomial function with $M\ll d$, then the latter parametrization may be favorable for minimizing the complexity. \\

In order to account for the dependence of the complexity on the underlying Pfaffian chain, it would be desirable to define an optimal Pfaffian complexity that can be assigned to a function. Roughly speaking, this should amount to finding the most optimal way of formulating a function in the Pfaffian framework, and could, for example, implement a minimization of the complexity over all Pfaffian chains. However, since the Pfaffian complexity of a function is measured by four integers, there is no unique choice to do so. Already for simple examples, such as the fewnomials discussed above, it can be subtle to find out which Pfaffian chain is optimal. However, as we will see in the next subsection, the numbers appearing in the Pfaffian complexity play a key role in the complexity of computations involving Pfaffian functions that we introduce in section~\ref{sec:complex_Pfaff}. A notion of optimal Pfaffian complexity could then be the one which minimizes this computational complexity. This observation will play a fundamental role later in section~\ref{sec:sharp-ominimal}.

When performing manipulations on Pfaffian functions, it is useful to keep in mind that the class of Pfaffian functions is closed under several operations, such as taking sums, products, derivatives, and compositions. In fact, it is possible to explicitly track the complexity of the resulting functions:
\begin{itemize}
\item If $f_1$ and $f_2$ are Pfaffian with complexity $\cC_{\zeta_i}(f_i)=(n,r_i,\alpha_i,\beta_i)$ for $i=1,2$, then the sum $f_1+f_2$ and the product $f_1f_2$ are Pfaffian with complexity
\begin{equation}
\label{ComplexityRules}
\begin{aligned}
\cC_{\zeta_1\cup\zeta_2}(f_1+f_2)&=(n,r_1+r_2,\max(\alpha_1,\alpha_2),\max(\beta_1,\beta_2)), \\
\cC_{\zeta_1\cup \zeta_2}(f_1f_2)&=(n,r_1+r_2,\max(\alpha_1,\alpha_2),\beta_1+\beta_2).
\end{aligned}
\end{equation}
Here $\zeta_1\cup\zeta_2$ denotes the Pfaffian chain formed by assembling the chains $\zeta_1$ and $\zeta_2$ into a single chain. Note that if there is overlap in the chains $\zeta_1$ and $\zeta_2$, then the order of the chain for the sum and product is less than $r_1+r_2$.

\item If $f$ is Pfaffian with complexity $\cC_\zeta(f)=(n,r,\alpha,\beta)$, then the partial derivative $\partial f/\partial x_j$ is Pfaffian in the same chain, with complexity 
\[
\cC_\zeta\left(\frac{\partial f}{\partial x_j}\right)=(n,r,\alpha,\alpha+\beta-1)
\]
\item If $f_1:U_1\to \bbR$ and $f_2:U_2\to \bbR$ are Pfaffian functions of complexity $\cC_{\zeta_i}(f_i)=(1,r_i,\alpha_i,\beta_i)$, where $U_1$ and $U_2$ are open intervals such that the image $f_1(U_1)$ is contained in $U_2$, 
then the composition $f_2\circ f_1$ is Pfaffian with complexity
\begin{equation}
\label{ComplexityRules2}
\cC_{\zeta_2\circ\zeta_1}(f_2\circ f_1) = (1,r_1+r_2,\alpha_2\beta_1 +\alpha_1+\beta_1-1,\beta_2).
\end{equation}
This can be shown by using the chain rule for partial derivatives. Here $\zeta_2\circ \zeta_1$ denotes a composite Pfaffian chain, which consists of the chain $\zeta_1$ together with the composite of $f_1$ with all the functions in the chain $\zeta_2$.
\end{itemize}

We will refer to these rules several times in the remainder of this work.

\subsubsection*{Pfaffian structures and tameness}
We are now able to introduce the structures of interest for this work. Let $\zeta$ be a given Pfaffian chain. Following the procedure of the previous subsection, we may consider the structure $\bbR_\zeta$ which is generated by the functions $\zeta_1,\ldots,\zeta_r$ in the chain. We will call such a structure a Pfaffian structure. Note that the Pfaffian structure $\bbR_\zeta$ automatically includes all Pfaffian functions $f(x_1,\ldots,x_n) = P(x_1,\ldots,x_n,\zeta_1,\ldots,\zeta_r)$ built from this chain, since polynomials are included in any structure.

Taking a step further, one may consider the set of \textit{all} Pfaffian functions, i.e.~all functions for which there exists an underlying Pfaffian chain. This collection of functions generates an even larger structure denoted by $\bbR_\text{Pfaff}$. In fact, since any Pfaffian chain is contained in this collection, all Pfaffian structures reside inside this larger structure. As a non-trivial mathematical result, Wilkie has shown that the structure $\bbR_\text{Pfaff}$ is o-minimal \cite{WilkieRPfaff}, and consequently for any Pfaffian chain $\zeta$ the smaller structure $\bbR_\zeta$ is o-minimal as well. This means that geometric objects built from Pfaffian chains are always tame.

It is worth pointing out that the triangularity condition in a Pfaffian chain is crucial for tameness. To illustrate this, consider the system of equations
\begin{align}
\frac{\partial \zeta_1}{\partial x} &=   \zeta_2, \\
\frac{\partial \zeta_2}{\partial x} & = -\zeta_1, \nonumber
\end{align}
which is not a Pfaffian chain since it is not triangular. The solutions to this system include the cosine function defined on the whole real line, which is not tame as it cannot be defined in any o-minimal structure. The cosine function is only tame when restricted to a finite interval.

Note that this can also be seen from Example (iv) above, where we constructed an explicit Pfaffian chain for the cosine function on the bounded interval $(-\pi,\pi)$. In fact, we saw that the function $x\mapsto \cos(mx)$ has a complexity whose degree scales with $m$. This allows us to relate complexity to mode expansions of periodic functions. Let us consider a periodic function $f$ on $(-\pi,\pi)$, and assume it has a finite Fourier expansion of the form 
\begin{equation}
f(x) = \sum_{m=0}^N (a_m\cos(mx)+ b_m \sin (mx) ).  
\end{equation}
The complexity of this function is determined by the highest frequency mode that is included in the sum. 

While section 3 is mainly devoted to physical examples, it is worth noting already that expansions of this form appear abundantly in physics. For instance, they appear as waves on a compact space, vibrational modes on a string, or Kaluza-Klein modes in a circle compactification. The discussion above shows that the complexity of such a mode is proportional to its frequency.

\subsection{Complexity in Pfaffian structures} 
\label{sec:complex_Pfaff}
In the previous section we have assigned a complexity to Pfaffian functions, and below we will shed light on the meaning of this terminology. The objects of interest are geometric objects built from Pfaffian functions, and one can assign various established notions of complexity to these objects, such as topological complexity and computational complexity. There exist several powerful theorems which show that there exist bounds on these quantities. Crucially, these bounds depend on the Pfaffian complexities of the underlying Pfaffian functions, which justifies the term `complexity'. In the following subsection we give a brief survey of these complexity theorems.

\subsubsection*{Complexity of Pfaffian equations}
The first geometric object that we will consider is the set of solutions to a system of equations involving Pfaffian functions. As a prelude, consider a set of $n$-variable polynomials $P_1,\ldots,P_n$ of degrees $\beta_1,\ldots,\beta_n$, respectively. The zero sets of each $P_i$ defines a hypersurface in $\bbR^n$, and the intersection of all these hyperplanes is the solution set to the system of equations
\begin{equation}
P_1=\ldots = P_n =0.
\end{equation}
Assuming that there are no degeneracies, this is a finite set. Bézout's theorem, a classic theorem in algebraic geometry, states that the number of solutions to this system of equations is bounded by the product of the degrees of the polynomials, i.e. 
\begin{equation}
\# \{x\in \bbR^n \,|\,P_1(x)=\ldots = P_n(x) =0 \}  \leq  \beta_1 \cdots \beta_n.
\end{equation}

This theorem has been generalized to the setting of Pfaffian functions by Khovanskii \cite{Khovanskii}. Consider a Pfaffian chain $\zeta$ on an open box $U\subseteq \bbR^n$, together with a set of Pfaffian functions $f_1,\ldots,f_n$ with complexity $\cC(f_i)=(n,r,\alpha,\beta_i)$ for $i=1,\ldots,n$. We now consider the set of solutions to the system of equations
\begin{equation} \label{f=0system}
f_1 = \ldots = f_n = 0\, ,
\end{equation}
Khovanskii's theorem now says that the number of non-degenerate solutions\footnote{By non-degenerate, we mean that the Jacobian determinant $\det(\partial f_i/ \partial x_j)$ is non-zero.} to this equation is bounded by 
\begin{equation}
2^{r(r-1)/2} \beta_1 \cdots \beta_n \big( \text{min}(n,r)\alpha + \beta_1 + \cdots + \beta_n -n+1\big)^{r}.
\end{equation}
In addition to the factor $\beta_1\cdots\beta_n$ which appears in Bézout's theorem, there are now several factors growing exponentially in the order $r$ of the chain. The number of intersection points of the hypersurfaces $\{f_i=0\}$ gives a rough measure of the complexity of this system of equations, and this theorem shows that this complexity has a bound determined by the complexities of the Pfaffian functions. 

It is worth noting that the o-minimality of Pfaffian functions already implies that the solution set of \eqref{f=0system} is finite. The fact that we work in a specific structure, namely the Pfaffian o-minimal structure $\bbR_\text{Pfaff}$, allows us to give an explicit bound.

\subsubsection*{Topological complexity}
The theorem discussed above extends to a much larger setting. Instead of focusing on systems of equations where the number of functions and variables is equal, we consider sets defined by an arbitrary number of equations. In addition, we now also allow for sets defined by inequalities. An \textit{elementary semi-Pfaffian set} is a set in $\bbR^n$ which is defined by solutions to equations and inequalities of Pfaffian functions. More precisely, it is a set of the form
\begin{equation}
\{x\in U \, | \, f_1(x) =0, \,\ldots,\,f_I(x)=0, \,g_1(x)>0, \ldots,\, g_J(x) >0  \},
\end{equation}
where the $f_i$ and $g_j$ are Pfaffian functions with a common Pfaffian chain defined on the domain $U$. A general semi-Pfaffian set is a finite union of elementary semi-Pfaffian sets, meaning that it can be written as 
\begin{equation} \label{def-Xsemi}
    X = \bigcup_{1\leq i\leq M}\{x\in U\,|\,f_{i1}(x) =0, \,\ldots,\,f_{iI_i}(x)=0, \,g_{i1}(x)>0, \ldots,\, g_{iJ_i}(x) >0  \}\ .
\end{equation}

In what follows we will be interested in the complexity of semi-Pfaffian sets. Given a semi-Pfaffian set $X\subseteq \bbR^n$, a good way to measure its \textit{topological complexity} is to look at the sum of the Betti numbers, i.e.~to consider 
\begin{equation}
b(X) = b_0(X) + \ldots + b_n(X),
\end{equation}
where $b_i(X) = \text{dim}(H_i(X;\bbR))$. Suppose that $X$ is a semi-Pfaffian set defined using $M$ equalities or inequalities of Pfaffian functions as in \eqref{def-Xsemi}, with each function having complexity $(n,r,\alpha,\beta)$ and having the same underlying Pfaffian chain. Then, as a generalization of Khovanskii's theorem, it was shown in \cite{GabVor04} that there is a bound on the topological complexity given by 
\begin{equation}
    b(X) \leq M^2 2^{r(r-1)/2} O(\text{min}(n,r)\alpha + n\beta )^{n+r}. 
\end{equation}
It is interesting to note how the various components of the complexity appear in this bound. It depends only polynomially on the degrees $\alpha$ and $\beta$, but exponentially on the number of variables $n$ and the length of the chain $r$.

\subsubsection*{Computational complexity of cell decomposition}
As mentioned above, the cell decomposition theorem says that every tame set admits a decomposition into finitely many cells. Since every tame set can be constructed from these elementary cells, it is natural to ask if there exists an algorithm which explicitly finds the cells needed to construct a given tame set.\footnote{This is reminiscent of quantum computing, where one of the main goals is to construct a given unitary operator from an elementary set of gates.} Whereas for general tame sets such an algorithm is not available, there does exist an explicit algorithm for semi-Pfaffian sets \cite{GabVor01}. There are various interesting quantities associated to such an algorithm, and for each of them it is possible to estimate the growth as a function of the complexity of the underlying Pfaffian functions. To illustrate this, let $X\subseteq \bbR^n$ be a semi-Pfaffian set of dimension $d$, defined using $M$ equalities and inequalities of Pfaffian functions as in \eqref{def-Xsemi}, all of which have complexity equal to $(n,r,\alpha,\beta)$. In this setting, the algorithm which finds the cell decomposition of $X$ has a computational complexity given by
\begin{equation} \label{CompComplex}
M^{(r+n)^{O(d)}}(\alpha + \beta) ^{(r+n)^{O(d^2n)}}.
\end{equation}
The dimension $d$ of the semi-Pfaffian set plays a key role in this estimate. The degrees $\alpha$ and $\beta$ again appear polynomially, whereas the number of variables $n$ appear double exponentially. For more details on the computational complexity of these algorithms we refer to \cite{GabVor01}.

%%%%%%%%%%%%%%%%%%%%%%%%%%
%%%%%%%%%%%%%%%%%%%%%%%%%%
\section{Pfaffian complexity in quantum systems}\label{sec:quantumsystems}

In this section we discuss how the notion of complexity introduced in section~\ref{Pfaff_definability} can be attached to physical quantities in various quantum systems. To accomplish this, we explicitly identify Pfaffian functions in three rather different settings: (1) correlation functions in simple QFTs evaluated on a point; (2) correlation functions and wavefunctions in quantum mechanics; (3) coupling functions in Seiberg-Witten theory. An important common feature of these setups is that we are able to analyze exact, non-perturbative physical quantities. Let us emphasize that the fact that these theories admit a tame structure is in itself already remarkable and in accordance with the observations of \cite{Grimm:2021vpn,Douglas:2022ynw,Douglas:2023fcg}. Here we find that this structure is actually a Pfaffian structure and hence allows us to assign a complexity to the arising functions. We find that, despite its appearance in these settings, the Pfaffian framework exhibits some limitations, which motivates us to extend our discussion to the more general framework of sharp o-minimality in sections \ref{sec:sharp-ominimal} and \ref{sec:sharp_physics}.

\subsection{QFTs on a point} \label{QFTsonPoint}

In this section we study zero-dimensional QFTs, and in particular those for which the spacetime consists of a single point. In these rather simplistic settings we are able to evaluate the partition function and correlation functions explicitly at a non-perturbative level and study their behaviour as a function of the parameters of the theory. The tameness of the partition functions of these theories was already discussed in \cite{Douglas:2022ynw}.\footnote{In the language of \cite{Douglas:2023fcg}, we investigate the tameness of the structure $\bbR_{\cT,\cS}$ which captures the observables of a set of theories $\cT$ formulated on a collection of spacetimes $\cS$. Here $\cT$ is parametrized by the couplings of the theory, and $\cS$ only contains a one-point spacetime.}

For simplicity we will consider a QFT with a single scalar field $\phi$, whose dynamics is encoded by an action $S$. Since the QFT is defined at a single point, the partition function and the correlation functions are ordinary integrals of the general form 
\begin{equation}
    I_{\nu}=\int_{-\infty}^{\infty}{\mathrm{d}}\phi \,\phi^\nu e^{-S}\, .
\end{equation}
These functions are related among each other through integration-by-parts identities. Using these relations it is possible to show that there exists a finite basis of integrals, such that every correlation can be expressed as a linear combination of the basis integrals. More precisely, if $J$ is the power of the highest-order interaction in the theory, then the set of correlation functions is spanned by the functions $I_0,\ldots,I_{J-2}$ with coefficients given by Laurent polynomials in the couplings of the theory. This idea was recently exploited to set up a system of differential equations satisfied by the basis correlation functions \cite{Weinzierl:2022mmp}.

\subsubsection*{Example 1: $\phi^4$-theory on one point}
We first focus on a massive scalar field with a single $\phi^4$-interaction. The Euclidean action of this theory, in the standard convention, is given by
\begin{equation} \label{phi4action}
S = \frac{m^2}{2}\phi^2 + \frac{\lambda}{4!}\phi^4 \,. 
\end{equation}
The parameters of the theory are the mass $m$ and the coupling $\lambda$, but since one is free to rescale the field $\phi$, the theory can be fully specified by a single parameter. In fact, reparametrizing the theory may also lead to a reduction in the complexity of computable quantities. This motivates us to rescale the field $\phi$ and introduce a new coupling parameter as follows
\begin{equation*}
    \phi\rightarrow\sqrt{\frac{3}{2 \lambda}} m \phi, \qquad g= \frac{3m^4}{4\lambda} \,.
\end{equation*}
In these variables, the action takes the simpler form
\begin{equation}
    S=g \phi^2 +\frac{g}{8} \phi^4 \,,
\end{equation}
and the coupling $g$ is now an overall multiplicative parameter. The general theory mentioned above tells us that a spanning set of correlation functions is given by $\{I_0,I_1,I_2\}$. However, since $I_1=0$ by symmetry, the integrals $I_0$ and $I_2$ suffice. The explicit form of these basis integrals is  
\begin{equation}
     I_0=\sqrt{2} e^{g} K_{1/4}(g),\qquad  
    I_2=-2\sqrt{2} e^{g}(K_{1/4}(g)-K_{3/4}(g)) \,,
\end{equation}
where $K_\alpha$ is the modified Bessel function of the second kind.

The correlation functions are related by an integration-by-parts (IBP) identity
\begin{equation}\label{eq:IBPpoint}
    I_\nu = \frac{2(\nu-3)}{g}I_{\nu-4}-4 I_{\nu-2} \,. 
\end{equation}
By iteratively invoking this identity, it is possible to write any $I_{\nu}$ as a linear combination of the two basis functions $I_0$ and $I_2$, with coefficients given by polynomials in $1/g$. The degree of these polynomials is determined by the amount of times the IBP identity has to be used. For example, we have
\begin{align*}
    I_4 &= \frac{6}{g}I_0 -4I_2\, ,\\ 
    I_6 &= -\frac{24}{g}I_0 + \left(16 +\frac{18}{g} \right)I_2\, , \\
    I_8 & = \left(\frac{96}{g} +\frac{180}{g^2} \right)I_0 + \left(-64 -\frac{192}{g} \right)I_2\, .
\end{align*}
In general, the degree of the polynomials appearing in the expansion of $I_\nu$ is bounded above by $\nu/4$.

We now claim that all the correlation functions $I_\nu$ are Pfaffian, enabling us to compute their complexity. To show this, we construct an explicit Pfaffian chain for $I_0$ and $I_2$. First, note that the derivatives of the correlators are 
\begin{equation}
\label{derivative}
   \frac{\pd I_\nu}{\pd g }= -I_{\nu+2}-\frac{1}{8} I_{\nu+4}\,.
\end{equation}
Together with the IBP relation of equation \eqref{eq:IBPpoint} this allows us to write down a second-order differential equation for $I_0$, namely
\begin{equation}
\label{phi4equation}
    \frac{\pd^2 I_0}{\pd g^2} -\left(2-\frac{1}{g}\right)\frac{\pd I_0}{\pd g}-\left(\frac{1}{g} + \frac{1}{16g^2}\right)I_0 =0 \,.
\end{equation}
Making use of the fact that $I_0$ is non-vanishing, it is possible to reduce this to a first-order equation by introducing the auxiliary function
\begin{equation}
    h(g) = -\frac{1}{I_0} \frac{\pd I_0}{\pd g} \,,
\end{equation}
which fulfills the so-called Riccati equation
\begin{equation}
  \frac{\pd h}{\pd g}=h(g)^2+\left(2-\frac{1}{g}\right) h(g)-\frac{1}{16 g^2}-\frac{1}{g}.  
\end{equation}
These functions fit together into a Pfaffian chain as follows
\begin{align}
    \zeta_1(g)& = \frac{1}{g}  &&  \frac{\pd\zeta_1}{\pd g}=- \zeta_1^2 \,, \\
    \zeta_2(g)&=h(g)   && \frac{\pd\zeta_2}{\pd g}= \zeta_2^2+\left(2-\zeta_1\right) \zeta_2-\frac{1}{16 }\zeta_1^2-\zeta_1\,, \nonumber\\ 
    \zeta_3(g)&=I_0(g) && \frac{\pd\zeta_3 }{\pd g} =-\zeta_2\zeta_3 \,. \nonumber
\end{align}

Within this chain, $I_2$ can be expressed as a Pfaffian function  
\begin{equation}\label{eq:I2Pfaff}
    I_2=2\zeta_2\zeta_3-\frac{1}{2}\zeta_1\zeta_3\, , 
\end{equation}
which follows from the IBP relation and the derivative relation in equation \eqref{derivative}. Since any correlation function $I_\nu$ can be written as a linear combination of $I_0$ and $I_2$, with coefficients given by polynomials in $\zeta_1$, it follows that all correlation functions in this theory are Pfaffian functions. 

The Pfaffian complexity of the correlation functions are now determined by this Pfaffian chain. First, consider the partition function $Z(g)=I_0(g)$ of the theory. Its complexity\footnote{For the convenience of the reader we recall that the complexity is defined to be the 4-tuple $(n,r,\alpha,\beta)$, where $n$ is the number of variables, $r$ is the order (or length) of the chain, and $\alpha$ and $\beta$ are the degrees of the polynomials involved in the construction.} can be read off from the chain as
\begin{equation}
    \cC(Z) = (1,3,2,1)\, .
\end{equation}
Next, the correlator $I_2$ is a Pfaffian function of degree 2 with respect to the given chain as shown by equation \eqref{eq:I2Pfaff}, so its Pfaffian complexity is $\cC(I_2) = (1,3,2,2)$. For general correlation functions, we use the fact that the expansion in terms of $I_0$ and $I_2$ has coefficients which are polynomial in the function $\zeta_1(g)=1/g$. From this we infer that the Pfaffian complexity of the correlation function $I_\nu$ (with $\nu$ assumed to be even, since it vanishes otherwise) is given by 
\begin{equation}
\cC(I_\nu) = \left(1,3,2, \left\lceil  \frac{\nu}{4} \right\rceil +1 \right) \qquad (\nu \text{  even})\, .
\end{equation}
Here $\lceil\cdot\rceil$ denotes the ceiling function. It is natural to expect that the Pfaffian complexity of the correlation function grows with the number of field insertions $\nu$, and our calculation shows that this growth is (stepwise) linear. Moreover, the degree $\beta$ is the only complexity parameter that grows, whereas the other parameters appear to be fixed by the theory.\footnote{This includes the parametrization of the theory; in a different description, the complexity could take a different value.}

\subsubsection*{Example 2: $\phi^6$-theory on one point}
A natural continuation is to study the $\phi^6$-theory on a point. For this theory, we choose the following parametrization of the action:
\begin{equation}
\label{eq:actionphi6}
    S=-\frac{g}{2}\phi^2+\frac{1}{96}\phi^6\;.
\end{equation}
The structure of the theory is similar to the $\phi^4$-theory, and the main difference is that the set of basis integrals now consists of the three functions $I_0$, $I_2$, and $I_4$. Note that, for technical reasons on which we will comment later, the mass term now carries a different sign than in the previous example. With this normalization, the IBP relation takes the form

\begin{equation}
   I_\nu= 16 \left((\nu -5) I_{\nu -6} +g I_{\nu -4}\right)\;. 
\end{equation}

The three basis correlation functions are explicitly given by
\begin{align*}
    I_0(g) &=\sqrt{2} \pi ^{3/2} \left(\text{Ai}(g)^2+\text{Bi}(g)^2\right) \,,\\
    I_2(g) &=4 \sqrt{2} \pi ^{3/2} (\text{Ai}(g) \text{Ai}'(g)+\text{Bi}(g) \text{Bi}'(g)) \,, \\
    I_4(g) &=8 \sqrt{2} \pi ^{3/2} \left(\text{Ai}'(g)^2+\text{Bi}'(g)^2+g \left(\text{Ai}(g)^2+\text{Bi}(g)^2\right)\right) \,,
\end{align*}
where we use a prime to note the first derivative, and where Ai and Bi are the Airy functions fulfilling the Airy equation
\begin{equation}
    \frac{\pd^2 f }{ \pd x^2}- x f(x)=0\;.
\end{equation}
We now claim that $I_0$, $I_2$ and $I_4$ are Pfaffian, and our strategy to show this will be to argue that the Airy functions are Pfaffian. To this end, we introduce the logarithmic derivatives $h_A=\text{Ai}'/\text{Ai}$ and $h_B=\text{Bi}'/\text{Bi}$ which satisfy
\begin{equation}
    \frac{\pd h_A }{\pd g}=h_A^2+g\,, \quad \frac{\pd h_B }{\pd g}=h_B^2+g\,.
\end{equation}
Note that these are only defined on a domain on which the Airy functions are non-vanishing. Since these are first-order, we are now able to write down the following Pfaffian chain: 
\begin{align}
    \zeta_1(g) &= h_A(g)   && \frac{\pd\zeta_1}{\pd g}= \zeta_1^2 +g \,, \\ 
    \zeta_2(g) &= \text{Ai}(g) && \frac{\pd\zeta_2 }{\pd g} =\zeta_1\zeta_2 \,, \nonumber \\
    \zeta_3(g) &= h_B(g)   && \frac{\pd\zeta_3}{\pd g}= \zeta_3^2 +g \,, \nonumber\\ 
    \zeta_4(g) &= \text{Bi}(g) && \frac{\pd\zeta_4 }{\pd g} =\zeta_3\zeta_4 \,. \nonumber
\end{align}

It follows that the Airy functions Ai$(g)$ and Bi$(g)$ are Pfaffian on a domain in which the logarithmic derivatives can be defined. Since the Airy functions are non-vanishing for $g>0$ and have infinitely many zeros\footnote{It is interesting to note that the presence of infinitely many zeros means that the Airy functions are not tame on the real line, but that the combination appearing in the correlation functions $I_0$, $I_2$, and $I_4$ is.} for $g<0$, we restrict to positive $g$. This explains the choice of the sign of the mass term in the action in equation \eqref{eq:actionphi6}.

Having set up the Pfaffian chain for the Airy functions, let us calculate the Pfaffian complexity of the correlation functions, starting with the three basis correlators. Firstly, the partition function $Z(g)=I_0(g)$ is quadratic in $\zeta_1$, so its Pfaffian complexity is 
\begin{equation}
    \cC(Z) = (1,4,2,2).
\end{equation}
Recalling that the partition function of the $\phi^4$-theory had Pfaffian complexity $(1,3,2,1)$, we can already observe that the complexity has increased as a consequence of the higher order interaction term. The correlators $I_2$ and $I_4$ involve the derivatives of the Airy functions, which can be written with degree $2$ as $\zeta_1\zeta_2$ and $\zeta_3 \zeta_4$, respectively. It follows that 
\begin{equation}
    \cC(I_2) = (1,4,2,3 ) , \quad  \cC(I_4) = (1,4,2,4) \,.
\end{equation}
For the higher correlation functions $I_\nu$, we use that they can be written as a linear combination of $I_0$, $I_2$ and $I_4$ with polynomial coefficients in $g$. As in the $\phi^4$-theory, the degrees of these polynomials depend linearly on $\nu$. From this we infer that, for $\nu$ even, 
\begin{equation}
\cC(I_\nu) = \left(1,4,2, \left\lfloor \frac{\nu}{4} \right\rfloor+3 \right) \qquad (\nu \text{  even and  } \nu\neq  0)\, ,
\end{equation}
where $\lfloor\cdot\rfloor$ denotes the floor function. The degree of the Pfaffian complexity again grows linearly with the number of field insertions $\nu$. It is interesting to note that the growth of the degree has the same rate as the complexity of the correlation functions in $\phi^4$-theory, whereas the `initial' value of the complexity is higher. Let us also comment on the fact that the order $r$ and degree $\alpha$ are independent of $\nu$. The reason for this is that all correlation functions can be expressed using the same Pfaffian chain, which is a remarkable consequence of the algebraic relations between the correlation functions. In general, one expects that more complicated physical quantities also require more involved Pfaffian chains.

Let us close with the observation that 
the functions appearing in the $\phi^4$- and $\phi^6$ theory have also been studied in the context of resurgence \cite{Dorigoni:2014hea}. Resurgence techniques provide a global perspective on the coupling space. The $\phi^4$ theory is the classic example with a computable partition function that 
admits a weak coupling expansion around $\lambda=0$ which is a \textit{transseries} rather than a convergent series \cite{PhysRevD.19.2370}. This implies that this function cannot be definable in the o-minimal structure $\bbR_{\rm an,exp}$. We have seen, however, that 
with real $\lambda$ and $m$, and the signs from \eqref{phi4action}, a Pfaffian structure is large  enough to describe this example.
In the $\phi^6$ case we have observed that the correlation functions only appear to be Pfaffian for $g>0$. If we rotate the coupling in the complex plane, we cross a Stokes line, which generates a monodromy action on the solution space. This leads to the addition of the oscillatory solutions on the other sheets. The Pfaffian structure is no longer rich enough to define this behaviour. However, let us allude to the fact that the framework of sharp o-minimality, introduced in section~\ref{sec:sharp-ominimal}, incorporates a much larger class of functions, in which the full coupling space of the $\phi^6$ theory can be covered.

\subsection{Quantum Mechanics} \label{QuantumMPfaff}

After a discussion of zero-dimensional quantum field theory, a natural next step is to consider the addition of one dimension, which leads us to one-dimensional quantum field theory, or quantum mechanics. Following the previous section, we will begin by investigating the tameness and complexity of correlation functions. We will then shift our perspective from quantum fields to quantum mechanics, and consider the propagator of the harmonic oscillator. This propagator naturally depends on space, which inspires us to investigate the complexity of the spatial dependence of quantum mechanical quantities more generally. We will do this by calculating the complexity of the wavefunction in several examples. 

\subsubsection*{Correlation functions}
We consider a single free quantum field $x$ with mass $m$, living on a spacetime consisting of a single Euclidean time dimension. As the notation suggests, this field may equivalently be interpreted as the position of a quantum mechanical particle. The theory is thus described by the harmonic oscillator Hamiltonian. As in the previous subsection, we rescale the physical coupling so that the Hamiltonian takes the form 
\begin{equation}
    H =  \frac{1}{2} p^2 + \frac{1}{2} g^2 x^2 \,.
\end{equation}
Note that the coupling $g$ is proportional to the energy scale of the oscillator. The theory can be solved exactly by standard techniques. In the Heisenberg picture, the field $x$ depends on time, and the 2-point correlation function with respect to the vacuum can be computed to be 
\begin{equation}
   I_2(g; t_2,t_1) =  \bra{0}x(t_2) x(t_1)  \ket{0} = \frac{2}{g} e^{-g (t_2-t_1)} \,.
\end{equation}
With respect to the coupling $g$, this function fits into the following Pfaffian chain:
\begin{align}
    \zeta_1(g) &= \frac{1}{g}   && \frac{\pd\zeta_1}{\pd g}= -\zeta_1^2 \,, \\ 
    \zeta_2(g) &= e^{-g(t_2-t_1)} && \frac{\pd\zeta_2 }{\pd g} =-(t_2-t_1)\zeta_2 \,, \nonumber 
\end{align}
so that $I_2 = 2\zeta_1\zeta_2$. The Pfaffian complexity of $I_2$ is then given by
\begin{equation}
    \cC(I_2) = (1,2,2,2).
\end{equation}
In the free theory, higher-point correlation functions may be obtained using Wick's theorem. For instance, the four-point function is given by 
\begin{align}
    I_4(g; t_4,t_3,t_2,t_1) &=  \bra{0}x(t_4) x(t_3) x(t_2) x(t_1)  \ket{0} \\
    &= \frac{4}{g^2}\Big( 2e^{-g (t_4-t_3+t_2-t_1)} + e^{-g (t_4+t_3-t_2-t_1)}    \Big)  \,.
\end{align}
As a consequence of summing over possible pairings, two distinct terms appear. Generically, both of these terms must be added to the Pfaffian chain separately. Using a similar Pfaffian chain as the one above, it follows that the complexity of the 4-point function is 
\begin{equation}
    \cC(I_4) = (1,3,2,3) \, .
\end{equation}
Let us now consider a general $\nu$-point function, assuming that $\nu$ is even. By Wick's theorem, this correlation function will consist of a prefactor $1/g^{\nu/2}$ multiplying a sum of exponential terms. The number of distinct exponential terms is determined by the combinatorial problem of distributing plus and minus signs over the times $t_{\nu-1},\ldots,t_2$ of the operator insertions, and is given by 
\begin{equation}
   { {\nu-1} \choose {\frac{\nu-1}{2}}  }\,.
\end{equation}
Each of these terms must be added to the Pfaffian chain separately, so we find that the order of the chain grows as $\nu!$ with the number of field insertions $\nu$. More precisely, the Pfaffian complexity of the $\nu$-point correlator is
\begin{equation}
    \cC(I_\nu) = \left(1,  { {\nu-1} \choose {\frac{\nu-1}{2}}  }, 3,\frac{\nu}{2} +1\right)\,.
\end{equation}
Let us compare this to the results found for QFTs on a point in the previous subsection. For the lattice correlation functions, the degree showed a similar linear growth. However, the order of the chain was fixed, whereas here it grows factorially with the number of field insertions. The essential reason for this is that the QFT on a single point eliminates the combinatorics of field insertions. 

\subsubsection*{Harmonic oscillator -- propagator}
Viewing the free 1d field theory as a harmonic oscillator, there is a more natural quantity to compute than the correlation functions, namely the propagator
\begin{equation}
G(g;t_2,t_1,x_2,x_1)  = \bra{x_2} e ^{-(t_2-t_1)H}\ket{x_1}.
\end{equation}
This propagator can be computed exactly by means of a path integral, and is given by the Mehler kernel

\begin{equation} \label{Mehler}
    G(g;t_2,t_1,x_2,x_1)  = \sqrt{ \frac{g}{2\pi \sinh(g \Delta t)} } \exp\left( \frac{-g\left( \cosh(g \Delta t )(x_1^2+x_2^2) -2x_1x_2\right)}{2\sinh(g \Delta t )} \right) \,,
\end{equation}
where $\Delta{t} = t_2-t_1$. 
The Pfaffian complexity of $G$ as a function of $g$ may be found by systematically applying the rules for manipulating Pfaffian functions explained in section~\ref{sec:pfaffian_chains}. As a result, one finds that there exists a Pfaffian chain for $G$ with complexity
\begin{equation}
    \cC(G) = (1,6,12,3).
\end{equation}
This example indicates that the Pfaffian complexity agrees with the intuitive idea of how complicated a function is. The Mehler kernel \eqref{Mehler} involves several non-trivial functions which are composed in an intricate way, causing the length of the chain to be $6$. This complexity is actually inherent to the physical quantity $G(g)$. While one might be tempted to find an alternative coordinate to $g$ to simplify the expression, one quickly realizes that this only leads to a reshuffling of complexity among the factors of $G(g)$ and that no significant reduction is possible.

\subsubsection*{Harmonic oscillator -- wavefunctions}
So far we have investigated physical quantities as a function of the couplings of the theory, and we regarded the dependence on position as a parameter. In the following we change our perspective and regard position as the variable of the functions while viewing the couplings as external parameters. As a starting point, let us consider the wavefunctions of the eigenstates of the harmonic oscillator. Their spatial dependence is encoded in the time-independent Schrödinger equation\footnote{Here we consider the complexity of the functional dependence on $x$, so we restore the physical parameters $m$ and $\omega$.}
\begin{equation}
    -\frac{1}{2m}\psi''(x) + \frac{1}{2}m\omega^2 x^2 \psi(x) - E\psi(x) =0. 
\end{equation}
Recall that the solutions take the form 
\begin{equation}
\psi_n(x) = \frac{1}{\sqrt{2^n n!}} \left( \frac{m\omega}{\pi}\right)^{1/4} e^{-m\omega x^2/2} H_n(\sqrt{m \omega }x)
\end{equation}
where $H_n$ is the $n$th Hermite polynomial. The state with wavefunction $\psi_n$ has an energy given by $E_n = (n+\tfrac{1}{2})\omega $. Let us now show how the functions $\{\psi_n\}$ fit into a Pfaffian structure. Consider the following Pfaffian chain of length one: 
\begin{align}
    \zeta(x) &= e^{-m\omega  x^2/2}   && \frac{\pd\zeta}{\pd x}=  -m\omega \,x\,\zeta \,.
\end{align}
This gives us a Pfaffian chain containing the Gaussian function, which has order 1 and degree 2. The Hermite polynomial $H_n$ has degree $n$, so it follows that, for each $n$, the function $\psi_n = P_{n+1}(\zeta,x)$ is a degree $n+1$ polynomial in $x,\zeta$ and hence a Pfaffian function. Hence, the complexity of $\psi_n$ in this chain is given by 
\begin{equation}
    \cC(\psi_n ) =  (1,1,2,n+1).
\end{equation}
The function $\psi_n$ describes a state $\ket{n}$ in the theory, and we may now define the complexity of the state $\ket{n}$ to be the complexity of the underlying wavefunction $\psi_n$. This leads us to the simple observation that the complexity scales with the energy of an eigenstate.

\subsubsection*{General wavefunctions}

Let us now look at the more general setting and consider 
a one-dimensional quantum system on an interval $(a,b)$ 
that is constrained by a general potential $V(x)$. 
We assume that $V(x)$ is a polynomial of degree $d\geq 2$. The eigenstate wavefunctions now satisfy the general one-dimensional Schrödinger equation
\begin{equation} \label{Schro-gen}
    -\frac{1}{2m} \psi''(x) + \big(V(x)-E\big)\psi(x) =0\, .
\end{equation}
Differential 
equations of this type are special types of Sturm–Liouville equations and have been analyzed in detail, for example, in reference \cite{Teschl}. The solutions admit a discrete energy spectrum $E_0 < E_1 < E_2 < ...$, with a finite lowest energy $E_0$ corresponding to the ground state wavefunction $\psi_0$.

Since this is a second order differential equation, it admits a possible reduction to the Riccati equation, by means of the substitution $f= \psi'/\psi$. This substitution is only valid for non-vanishing wavefunctions $\psi$.  
According to the node theorem in quantum mechanics, the wavefunction of the $n$th excited state with energy $E_n$ has exactly $n$ zeros \cite{Teschl}. In particular, the ground state wavefunction $\psi_0(x)$ is non-vanishing. Therefore, the substitution $f_0=\psi'_0/\psi_0$ is well-defined and satisfies the Riccati equation
\begin{equation}
    \zeta_0'+\zeta_0^2 -2m\big(V(x) - E_0\big) =0.
\end{equation}
It follows that $\psi_0$ fits into the Pfaffian chain given by

\begin{align}
    \zeta_1(x) &= \frac{\psi'_0(x)}{\psi_0(x)}   && \frac{\pd\zeta_1}{\pd x}=  -\zeta_0^2+2m\big(V(x) - E_0 \big) \,, \\ 
    \zeta_2(x) &= \psi_0(x) && \frac{\pd\zeta_2 }{\pd x} =\zeta_1\zeta_2 \,, \nonumber 
\end{align}
which has order $2$ and degree $d$ specified by the degree of the potential $V(x)$. The Pfaffian complexity of the ground state with respect to this chain is now 
\begin{equation}
    \cC(\psi_0) = (1,2,d,1). 
\end{equation}
However, recalling the discussion of fewnomials from the previous section, we note that if $V$ only has a small number of monomial terms $M$ compared to the degree $d$, there exists a Pfaffian chain with a lower complexity. In this chain we include each of the monomials of $V$, and the resulting Pfaffian complexity of the ground state wavefunction is 
\begin{equation}
    \cC(\psi_0) = (1,M+3,2,1) \,.
\end{equation}
In general we infer that, in the Pfaffian framework, more complicated potentials lead to ground states of higher complexity.

\subsubsection*{Pfaffian complexity and quantum computational complexity}
Within the context of quantum-mechanical states, there exists another well-known notion of complexity, namely quantum computational complexity (for a detailed review we refer to \cite{Chapman:2021jbh}), and it is natural to investigate how these two concepts are related. As we will see, the two notions of complexity differ in a number of fundamental aspects.

The main idea behind quantum computational complexity is to construct a given target quantum state from a reference state by acting with a unitary operator $U$, which is to be constructed as a product of fundamental unitary operators called gates. These gates represent physical operations which may be performed on the states, and form a finite set. The connection of the reference state to the target state can be thought of as a circuit with finitely many nodes.   The quantum computational complexity of this state is then defined to be the minimum number of gates one needs to multiply to construct $U$ and hence gives a measure of how complex the circuit is. This notion of complexity initially only captures quantum systems with finite-dimensional Hilbert spaces. However, generalizations have been proposed which replace a discrete sequence of gates by a continuous unitary evolution of the reference state. Subsequently, these ideas have been applied to the harmonic oscillator, with the aim of assigning a quantum computational complexity to states in quantum field theories \cite{Jefferson:2017sdb}.\footnote{Similar ideas were implemented in the setting of conformal field theories \cite{Chagnet:2021uvi}.}

In order to compare the two notions, we enumerate three essential features of quantum computational complexity.
\begin{itemize}
     \item[(i)] It is defined \textit{relative} to another quantum state, and without a fixed reference state there is no inherent notion of quantum computational complexity;
     \item[(ii)] Applications require careful generalizations from discrete to continuous complexity. Extending the framework in this way requires making additional choices, such as picking a distance measure on the space of unitary operators;
     \item[(iii)] It depends in a natural way on the parameters of the theory. For example, in \cite{Jefferson:2017sdb} it was shown that the complexity of the ground state $\psi(x)\sim e^{-m\omega x^2/2}$ of the harmonic oscillator relative to the ground state $\psi_0(x)\sim e^{-m\omega_0 x^2/2}$ of another harmonic oscillator depends on the frequency ratio $\omega/\omega_0$.
\end{itemize}
Let us now comment on the relation of these features to Pfaffian complexity, and point out a number of open problems in understanding this connection. 

To address point (i), we note that Pfaffian complexity measures the complexity of physical quantities in an \textit{absolute} sense, encoding the amount of logical information required to describe a physical quantity. Instead of considering the complexity of a state relative to another state, the Pfaffian framework is naturally more apt to describing the complexity of logical statements that involve both states. 
In particular, it is desirable to understand how a quantum circuit can be described inside the logical framework, since we can then determine the Pfaffian complexity of the whole circuit.
If one has a notion of gates that one is allowed to use, then one can try to minimize this Pfaffian complexity on the space of allowed circuits.\footnote{Note that we do not see that the Pfaffian setting gives a natural notion of `gate'. The set of gates is still a choice that depends on the system under consideration.} 
In the spirit of the example discussed above, one may for instance calculate the complexity of statements involving $\psi(x) = e^{-m\omega x^2/2}$, $\psi_0(x) = e^{-m \omega_0 x^2/2}$, and all intermediate states connecting them in a circuit. We leave making this precise as an interesting open problem for the future.

Turning to point (ii), we note that Pfaffian complexity is fundamentally a discrete quantity, which admits no immediate continuous generalization. In the context of quantum circuits, one attempt to implement this generalization could be to adopt the geometric perspective on circuit complexity, replacing the search for an optimal discrete circuit by the optimization of a path in an operator space. The curve traced out by this path is a set, which means that one can assign a Pfaffian complexity to it. It would be interesting to investigate if this quantity can serve as a measure of complexity of the associated circuit.

Finally, addressing point (iii), we stress that the Pfaffian framework describes the complexity of the functional dependence on a set of variables. Therefore, the Pfaffian complexity itself cannot depend on these variables, and only on external parameters. Following the example above, consider the Pfaffian complexity of the function $\psi(x)=e^{-m\omega x^2/2}$ with respect to a Pfaffian chain $\zeta_0$ containing the function $\psi_0(x)= e^{-m\omega_0 x^2/2}$. Since these are functions of the variable $x$, the complexity will be independent of $x$. However, one may inquire about the dependence on the frequencies $\omega$ and $\omega_0$. The function $\psi$ is only Pfaffian in this chain in the special instance that the ratio $\omega/\omega_0$ is an integer, so that we have the algebraic relation $\psi=\psi_0^{\omega/\omega_0}$. In this case the complexity of $\psi$ is given by $\cC_{\zeta_0}(\psi)=(1,1,2,\omega/\omega_0)$, and the dependence on physical parameters appears. Generically, $\omega/\omega_0$ is not an integer, and in order for the wavefunction $\psi$ to be Pfaffian, the chain must be extended, leading to an increase in the Pfaffian complexity independent of $\omega$ and $\omega_0$.

\subsection{Seiberg-Witten theory and elliptic integrals}\label{sec:SW}

In our third example, we study a four-dimensional supersymmetric quantum field theory, namely the SU$(2)$ Seiberg-Witten theory \cite{Seiberg:1994rs} (for a detailed review we refer to \cite{SW}). Our motivation to consider this theory is that various physical quantities can be computed non-perturbatively due to the $\mathcal{N}=2$ supersymmetry. Instead of the correlation functions, we will now consider the coupling function which encodes the coupling of the gauge fields to the scalars.\footnote{Following the notation of \cite{Douglas:2023fcg}, this function is part of the structure $\bbR^\text{def}_{\cT,\cS}$ which contains the functions required to define a set of quantum theories $\cT$ formulated on a set of spacetimes $\cS$. }

The starting point in constructing the SU$(2)$ theories is 
an $\mathcal{N}=2$ Yang-Mills Lagrangian for a vector multiplet containing the SU$(2)$ gauge field.\footnote{One can also allow for having a 
number of charged hypermultiplets. We will not consider this more general case in this discussion.} Additionally, this multiplet contains a number of complex scalars $\Phi$. 
The kinetic term of the vector fields is determined by the (complexified) gauge coupling function
\begin{equation}
    \tau_0=\frac{\theta_0}{2\pi}+\frac{4\pi i}{g^2_0}\,.
\end{equation}

In the vacuum the gauge symmetry is spontaneously broken from SU$(2)$ to U$(1)$ by a vacuum expectation value $a$ of the scalar fields. The possible choices of the scalar $a$ parameterize the Coulomb branch of the theory. To remove the redundancy $a\rightarrow -a$, we introduce another coordinate 
$u=\text{Tr}(\Phi^2)=2a^2$. At sufficiently low energies the effective theory becomes the theory of a single U$(1)$ $\mathcal{N}=2$ vector multiplet whose dynamics is encoded by an effective prepotential $F(a)$. The gauge coupling function in then reads 
\beq
   \tau(a)=\frac{\partial a_D}{\partial a}\;,\qquad a_D = \frac{\partial F}{\partial a}\, ,
\eeq
where we have also defined the dual scalar field $a_D$.
Remarkably, the effective prepotential $F(a)$ can be computed explicitly, including all perturbative and non-perturbative corrections, by geometrizing the problem through the introduction of an auxiliary elliptic curve $\cC$ -- the Seiberg-Witten curve. Physically, the weak coupling point is at $u =\infty$, while the theory is strongly coupled near the point $u=0$, where in the classical theory the SU$(2)$ gauge symmetry is restored. In the complete quantum-corrected moduli space, the special point $u=0$ is replaced by two strong coupling points $u=\pm \Lambda^2$, where $\Lambda$ is a dynamically generated quantum scale.

The low-energy effective theory is obtained by 
integrating out the massive components of the SU$(2)$-vector multiplet and depends on the energy scale $\Lambda$. Supersymmetry protects the effective gauge coupling function $\tau(u,\Lambda)$ from perturbative corrections at more than one loop, while it generally permits the presence of an infinite sum of non-perturbative corrections. It takes the general form
\begin{equation}
    \tau(u,\Lambda)= \tau_0 +\frac{2i}{\pi }\log(\frac{\sqrt{u}}{\Lambda})+\sum_{n=1}^\infty a_n\left(\frac{\Lambda}{\sqrt{u}}\right)^{4 n}\,,
\end{equation}
where the logarithmic term encodes the one-loop correction.
In the SU(2) theory, $\tau(u,\Lambda)$ can be explicitly evaluated. Absorbing $\Lambda$, the strong coupling singularities are at $u=\pm 1$. The quantum-corrected fields $(a,a_D)$ are computed to be \cite{SW}

\begin{equation} \label{def-a-aD-k}
    \begin{pmatrix}
    a\\
    a_D
    \end{pmatrix}
    =\begin{pmatrix}
   \frac{4}{\pi k}E(k)\\
   \frac{4}{i \pi}\frac{E(1-k)-K(1-k)}{\sqrt{k}}
    \end{pmatrix}\, ,\qquad k=\frac{2}{1+u}\ .
\end{equation}
Here the elliptic integrals $E(x)$ and $K(x)$ are special cases of  
the Gauss hypergeometric function $\,_2F_1(a,b,c,x)$, in particular one finds that\footnote{Throughout this paper we use the elliptic parameter as the argument of the complete elliptic integrals. This convention is compatible with the ones used in Mathematica.}
\beq
K(x)=\frac{2}{\pi}\;_2F_1\big(\tfrac{1}{2},\tfrac{1}{2},1,x\big)\;, \qquad 
E(x)=\frac{2}{\pi}\;_2F_1\big(-\tfrac{1}{2},\tfrac{1}{2},1,x \big)\;.
\eeq
Using the properties of the Gauss hypergeometric function, the gauge coupling function can be expressed as 
\begin{equation} \label{tau-K}
    \tau(k) =i\frac{K(1-k)}{K(k)}\,.
\end{equation}

We now want to show that, at least on a real slice of the field space, the real and imaginary parts of $\tau$ are Pfaffian functions. Hence, from now on we 
consider the real half-line
\beq \label{u-domain}
    u \in (1,\infty)\, ,
\eeq
and comment on possible generalizations later.
Using \eqref{def-a-aD-k} we infer that this implies that $0 < k<1$, and hence that we need to evaluate the hypergeometric functions $K(x)$ for real values $0<x<1$ (this describes the function $K(k)$ as well as $K(1-k)$). The key point is that on this domain, the function $K(x)$ is actually Pfaffian.

Since $K(x)$ is a hypergeometric function, it satisfies the differential equation\footnote{Note that all Gauss hypergeometric functions $\,_2F_1(a,b,c,x)$ satisfy a second order differential equation depending on $(a,b,c)$. We only need to consider the special case $K(x)$.}
\begin{equation}
    \frac{\partial^2 K(x)}{\partial x^2}=\frac{ K(x) }{4(1-x)x}+\frac{2 x-1}{(1-x)x} \,\frac{\partial K(x)}{\partial x}\,.
\end{equation}
This equation can be transformed into the Riccati equation, as we have done in the previous examples. Let us set 
\beq
    h(x)=\frac{1}{K(x)} \frac{\partial K(x)}{\partial x} \,, 
\eeq
which obeys the first-order differential equation
\begin{equation}
    \frac{\partial h(x)}{\partial x}=\frac{1}{4(1-x)x}+\frac{2 x-1}{(1-x)x}\, h(x)+ h(x)^2\,.
\end{equation}
This enables us to write down the following Pfaffian chain for $K(x)$:

\begin{align}
    \zeta_1(x) &= \frac{1}{x}   && \frac{\pd\zeta_1}{\pd x}= -\zeta_1^2  \,, \\ 
    \zeta_2(x) &= \frac{1}{1-x} && \frac{\pd\zeta_2 }{\pd x} =-\zeta_2^2 \,, \nonumber \\
    \zeta_3(x) &= h(x)   && \frac{\pd\zeta_3}{\pd x}= \frac{1}{4}\zeta_1\zeta_2+(2x-1)\zeta_1\zeta_2 \zeta_3+\zeta_3^2 \,, \nonumber\\ 
    \zeta_4(x) &= K(x) && \frac{\pd\zeta_4 }{\pd x} =\zeta_3\zeta_4 \,. \nonumber
\end{align}

Let us now compute the Pfaffian complexity of the gauge coupling function $\tau$, given in equation \eqref{tau-K}. In our chosen domain, $\tau$ is purely imaginary. Using the composition rules for Pfaffian complexity explained in section~\ref{sec:pfaffian_chains}, we find that the complexity of $\text{Im}\,\tau(k)$ is given by
\begin{equation} \label{Ctau-k}
\cC(\text{Im}\,\tau) = (1,7,4,2)\, ,
\end{equation}
where we evaluate $\tau$ as a function of the coordinate $k$.
A more natural coordinate in the theory is $u$, and as a function of $u$ the gauge coupling has complexity
\begin{equation} \label{Ctau-u}
    \cC(\text{Im}\,\tau) = (1,8,6,2)\,.
\end{equation}
This follows from applying the chain rule for Pfaffian complexity to the coordinate change from $k$ to $u$ as given in equation \eqref{def-a-aD-k}. An important observation is that the Pfaffian complexity, in either choice of coordinates, increases successively from the classical to the one-loop corrected to the fully non-perturbative expression~\eqref{tau-K}. Hence, even in our rather restricted analysis we can view the Pfaffian complexity as a measure of how many quantum contributions have been included.

Let us briefly comment on the fact that the change of coordinate from $k$ to $u$ leads to an increase in the complexity from \eqref{Ctau-k} to \eqref{Ctau-u}. That the Pfaffian complexity depends on the coordinate choice is expected and we have seen incarnations of this in section~\ref{sec:pfaffian_chains}. In fact, the Pfaffian complexity we computed is not purely intrinsic to the physical object but also reliant on the description of the object. We can always increase the complexity by choosing a more involved presentation, but the key point is that there are minimal presentations of the object and 
these will feature the minimal complexities. While a more clever choice of coordinates might reduce the Pfaffian complexity, it can only reduce it in a limited way if one wants to retain the physical properties of the objects.

It appears here that the coordinate $k$ allows one to describe the gauge coupling $\tau$ with less information, which seems to suggest that this is the correct coordinate to use. However, it is likely too naive to take this as a guide. In our analysis we focused on a single physical quantity in Seiberg-Witten theory, the gauge coupling function, in a real patch of the moduli space. We believe that a discussion of the coordinates leading to the presentation of the theory in which it has the smallest complexity should include all physical quantities of the theory and be more global in nature. Such an analysis goes beyond the scope of this work and would require to turn to the more general formalism introduced in section~\ref{sec:sharp-ominimal} as we discuss next.

An important next step is to generalize the discussion to the full 
quantum corrected field space parameterized by a complex $u$.
Here again, as in sections~\ref{QFTsonPoint} and  \ref{QuantumMPfaff} we reach the limitations of what is possible within Pfaffian structures. It is not expected that the periods of a Riemann surface, even for an elliptic curve, are Pfaffian functions. One obstruction is the appearance of monodromy symmetries arising when encircling the singularities in the complex $u$-field space. Similar to the discussion of resurgence at the end of section~\ref{QFTsonPoint}, we then have to deal with multiple sheets and how they are connected. Remarkably, this can indeed be done, but one needs to move to more general \textit{sharply} o-minimal structures. We will introduce this concept in the next section and then return to Seiberg-Witten theories in section~\ref{sec:sharp_physics}.

%%%%%%%%%%%%%%%%%%%%%%%%%%
%%%%%%%%%%%%%%%%%%%%%%%%%%

\section{Sharp o-minimality and complexity}
\label{sec:sharp-ominimal}
In the previous sections we have seen how it is possible to assign a complexity to certain tame objects built from Pfaffian chains, including several physical quantities naturally arising in quantum theories. However, in each of the studied examples we encountered limitations in the scope of the Pfaffian framework. The aim of this section is to introduce the concept of sharply o-minimal structures \cite{binyamini2022sharply}, which vastly generalizes the Pfaffian chain approach to tameness and complexity. Based on this framework, we define a new notion of complexity which we term $\sharp$complexity. Before giving the technical definitions, we start by motivating the idea of sharp o-minimality. After reviewing the basic definitions and properties of the framework, we discuss novel mathematical conjectures on sharp o-minimality which may shed light on the role of sharp o-minimality in physics. This section mostly takes a mathematical perspective, and the application to physics is deferred to section \ref{sec:sharp_physics}.

\subsection{Motivating sharp o-minimality}
Let us start by briefly recollecting what we have done so far. Our aim has been twofold. On the one hand we wish to capture the amount of information contained in physical quantities, in terms of the complexity of the functional dependence on physical parameters. On the other hand, we strive to gain a deeper understanding of tameness in physical theories. The framework of Pfaffian structures, with its notion of Pfaffian complexity, has proven to be a suitable setting to address both of these points. Let us summarize the main features of Pfaffian complexity:
\begin{itemize}
     \item[(i)] We consider the simplest way of describing a tame physical function, in terms of a Pfaffian system of differential equations; 
     \item[(ii)] The complexity of this description is encoded in a few integers;
     \item[(iii)] From these integers it is possible to compute bounds on the complexity of logical operations and computations involving these functions, justifying the name `complexity'.
\end{itemize}

The limitations of the Pfaffian framework mostly arise from point (i), since many tame physical quantities are not Pfaffian functions. As a consequence of tameness, these functions do have a finite complexity, and this indicates the need for a larger framework which includes all tame functions arising in physics. 
From a more technical point of view, another disadvantage of Pfaffian complexity lies in point (iii). When considering general Pfaffian sets described using logical statements involving Pfaffian functions, it turns out that Pfaffian complexity is not compatible with all elementary logical operations.\footnote{To be precise, technical challenges arise when using the logical operation of negation ($\neg)$, which geometrically corresponds to taking the complement of a set. }

Inspired by the foundational notion of Pfaffian complexity, and with the aim of addressing these limitations and generalizing the notion of complexity in tame geometry, Binyamini and Novikov have introduced the notion of \textit{sharp o-minimality} \cite{binyamini2022sharply,BinNovICM}. This is a refinement of o-minimality as introduced in section~\ref{sec:tameness}, based on an additional number of axioms. Essentially, these axioms are generalized and abstract versions of points (ii) and (iii) in the summary above.

Instead of the four numbers appearing in the Pfaffian complexity, sharp o-minimality is based on two integer numbers, called \textit{format} $F$ and \textit{degree} $D$. This reduction to two numbers is inspired by the fact that for the Pfaffian complexity $\cC(f)=(n,r,\alpha,\beta)$, the number of variables $n$ and the order $r$ (and respectively, the degrees $\alpha$ and $\beta$) appear on a similar footing. This is evident, for instance, in the formulas for the topological and computational complexity given in section~\ref{sec:complex_Pfaff}. These numbers $F$ and $D$ are implemented for all definable sets in a sharply o-minimal structure, making it far more widely applicable than Pfaffian complexity. Moreover, conditions are imposed on $(F,D)$ to ensure that they are consistent with \textit{all} logical operations needed in tame geometry. To axiomatize point (iii), the definition of sharply o-minimal structures includes a collection of polynomials $P_F(D)$, which encode complexity bounds on logical and computational operations involving definable sets. Crucially, these bounds are \textit{effective}, meaning that they can be computed explicitly. 

From a physical perspective, we will show in section~\ref{sec:sharp_physics} that sharp o-minimality allows us to, in principle, compute complexities of physical quantities which we were not able to address within the scope of the Pfaffian framework. This observation, combined with mathematical conjectures which roughly state that all tame geometric functions fit into a sharp o-minimal structure \cite{BinNovICM}, motivates us to think that sharp o-minimality is the ideal framework for describing complexity in tame physical systems. An additional advantage is that, whereas o-minimality only implies finiteness, sharp o-minimality provides computable bounds.

\subsection{Sharply o-minimal structures}
\subsubsection*{Format and degree}
Let us now turn to the technical definition of sharp o-minimality, starting with the notion of format $F$ and degree $D$. As pointed out in the motivational subsection, the purpose of these numbers is to quantify the complexity of all definable sets in $\cS$. Formally, this is done by introducing for every pair of positive integers $(F,D)$ a collection $\Omega_{F,D}$ which by definition contains all the definable sets of format and degree $(F,D)$. Together, these collections are assumed to form a filtration $\Omega={\Omega_{F,D}}$, meaning that
\begin{equation}
\Omega_{F,D} \subseteq \Omega_{F+1,D} \quad\text{and}\quad  \Omega_{F,D} \subseteq \Omega_{F,D+1}\,.
\end{equation}
Note that for a given definable set, the format and degree are therefore not uniquely defined; we will comment on this ambiguity later.\footnote{Note that this is equally true for Pfaffian complexity, since it is always possible to make the Pfaffian chain for a Pfaffian function more complicated.} Such a filtration $\Omega$ is called an FD-filtration.

\subsubsection*{Definition of sharply o-minimal structures}
In order for the FD-filtration $\Omega$ to describe a consistent and meaningful notion of complexity on an o-minimal structure $\cS$ we have to impose a number of additional conditions, and this then leads us to the definition of a \textit{sharply o-minimal structure}.\footnote{The adjective `sharp'  refers to the sharp complexity bounds which objects in such a structure satisfy.} The conditions are as follows:
\begin{itemize}
     \item[(i)] If $A \in \Omega_{F,D}$ and $A\subseteq \bbR^n$, then $F\geq n$; 
     \item[(ii)] If $A \in \Omega_{F,D}$ and $A\subseteq \bbR^n$, then 
     \begin{equation}
     \pi(A), \,\, \bbR^n\setminus A  \in \Omega_{F,D} \,,
     \end{equation}
     where $\pi$ is any linear projection $\bbR^n\to\bbR^{n-1}$, and
     \begin{equation}
     A\times \bbR, \,\,\bbR\times A \in\Omega_{F+1,D} \,.
     \end{equation}
     \item[(iii)] If $A_i \in \Omega_{F_i,D_i}$ for $i=1,\ldots,k$ and $A_i\subseteq \bbR^n$, then
     \begin{equation}
     \bigcup_{i=1}^ k A_i ,\,\,  \bigcap_{i=1}^ k A_i \in \Omega_{F,D} \,,
     \end{equation}
     where $F=\max_i\{ F_i\}$ and $D=\sum_i D_i$;
     \item[(iv)] If $P \in \bbR[x_1,\ldots,x_n]$ is a polynomial, then 
     \begin{equation}
     \{P= 0 \} \in \Omega_{n,\deg P} \,;
     \end{equation}
      \item[(v)] For every $F$ there is a polynomial with positive coefficients $P_F$ such that if $A \in \Omega_{F,D}$ and $A\subseteq \bbR$, then $A$ has at most $P_F(D)$ connected components.
\end{itemize}

The first axiom puts a lower bound on the format in terms of the ambient dimension of a definable set, which may be interpreted as a bound on the number of variables. Axioms (ii) and (iii) indicate how format and degree behave under elementary set-theoretical operations. In particular, it shows that for definable sets built by taking unions and intersections, the format $F$ serves as a bound on the formats of the constituent sets, whereas the degree is an additive quantity that grows when more sets are added. Then axiom (iv) constrains the degree and format associated to zero sets of polynomials. These sets are definable in any o-minimal structure and hence also in all sharply o-minimal structures. 
It shows that for algebraic sets, the degree is the expected notion, namely the degree of the underlying polynomial. Finally, the fifth axiom tells us how the format and degree of one-dimensional definable sets relate to the topology of these sets. This axiom is the crux of sharp o-minimality; whereas standard o-minimality merely places a finiteness condition on subsets of the real line, sharp o-minimality imposes an explicit upper bound, encoded in the polynomials $P_F$.  

\subsubsection*{Examples and non-examples} 

To gain some intuition for sharp o-minimality, let us discuss a two examples. First, recall from section \ref{sec:tameness} that the simplest o-minimal structure is $\bbR_\text{alg}$. The definable sets of this structure are generated by algebraic sets of the form $\{P=0\}$, where $P$ is a polynomial. Let us assign the format and degree $(F,D)=(n,\deg P)$ to these sets, where $n$ is the number of variables in $P$. The axioms (ii) and (iii) now determine the format and degree of all sets in $\bbR_\text{alg}$. This gives rise to an FD-filtration on $\bbR_\text{alg}$, and it was argued in \cite{binyamini2022sharply} that this makes $\bbR_\text{alg}$ into a sharply o-minimal structure. Even though $\bbR_\text{alg}$ is a relatively simple o-minimal structure, this result is non-trivial and relies on effective cell decomposition algorithms. The computational complexity of these algorithms encode the polynomials $P_F(D)$ needed in the definition of sharp o-minimality. 

The second example is a direct generalization of the framework of Pfaffian structures. Consider the o-minimal structure $\bbR_\text{rPfaff}$ generated by restricted Pfaffian functions. In reference \cite{BinVor22}, a notion of format and degree for the sets in $\bbR_\text{rPfaff}$ is defined in terms of the Pfaffian complexity of the underlying functions. Roughly speaking, the format of a Pfaffian function is given by $n+r$, and the degree is given by the maximum among the degrees $\alpha$ and $\beta$. For sets constructed from multiple Pfaffian functions, the format is the maximum of the underlying formats, and the degree is the sum of the underlying degrees. For the technical details of this construction we refer to \cite{binyamini2022sharply,BinVor22}. Note that this implies in particular that our computations from the previous section carry over to the more general framework of sharp complexity. 

To show that not every o-minimal structure is sharply o-minimal, let us note that $\bbR_{\rm an}$ is the prime example of a structure that cannot be sharply o-minimal. The precise argument 
can be found in \cite{binyamini2022sharply}, but it essentially uses the fact that for a general (restricted) analytic function one can choose too many free coefficients in its Taylor series. A clever choice can then be used to violate the polynomial growth required by sharp o-minimality. Clearly, this implies that also $\bbR_{\rm an, exp}$ cannot be sharply o-minimal. The fact that both $\bbR_{\rm an}$ and $\bbR_{\rm an,exp}$ are too large to be sharply o-minimal fits nicely with the intuition that a general analytic function contains too much information and can thus be arbitrarily complex.

\subsection{Sharp complexity -- $\sharp$complexity}
Let us now discuss the notion of complexity in the framework sharply o-minimal structures. As we alluded to earlier, the purpose of the format and degree is to quantify the geometric complexity of a tame set or function. The interpretation of this idea is that the format and degree control the various other notions of complexity arising when performing operations on these sets, such as topological complexity, computational complexity, and logical complexity. As an example, consider a definable set $X$ which is obtained by applying unions, intersections, projections, and complements to some definable sets $X_1,\ldots,X_k$ of format and degree $(F_j,D_j)$. One may then for instance consider the question of decomposing $X$ into connected components by running a geometric algorithm:\footnote{Note that the geometric problem of finding connected components may also be applied to the problem of finding the number of solutions of system of equations.} 
\[
\begin{tikzcd}[column sep=80pt]
X_1,\ldots,X_k \quad\arrow[r,"\text{geometric operations}"]&   X \arrow[r,"\text{algorithm}"]& 
\text{connected components of }X\ .
\end{tikzcd}
\]
The framework of sharp o-minimality is now designed such that all data associated to this operation, e.g.~the number of connected components of $X$, the complexity of the algorithm, or the amount of information needed to describe the connected components, are \textit{polynomially} dependent on the degrees $D_1,\ldots,D_k$, where the precise polynomial dependence is determined by the formats $F_1,\ldots,F_k$. This interpretation is made precise by several theorems pertaining to the fundamental geometric operations of tame geometry, such as cell decomposition \cite{binyamini2022sharply}.

\begin{figure}[h]
\begin{center}
\vspace*{.5cm}
 \includegraphics[width=.4\textwidth]{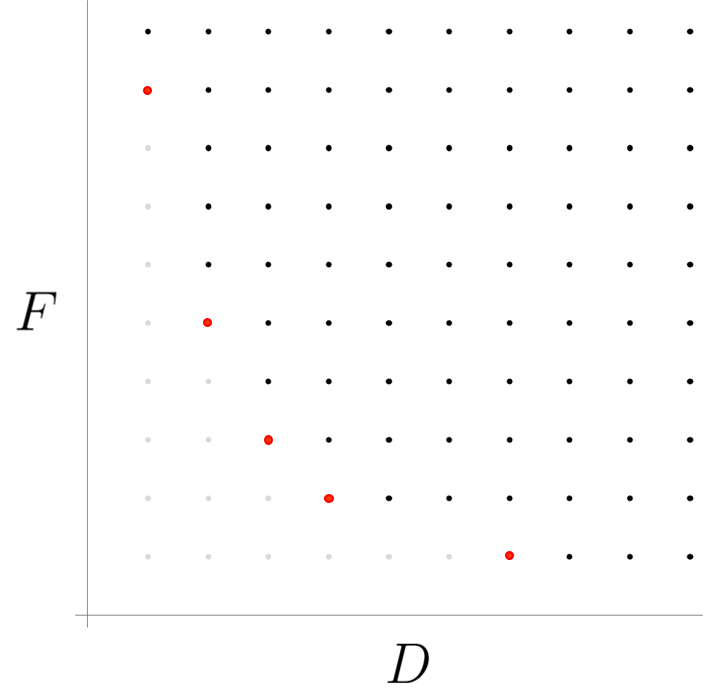} 
 \vspace*{.5cm}
\caption{An example of what the set of points $(F,D)$ for which $X\in\Omega_{F,D}$, represented by the black and red dots, may look like.
Together these define the complexity $\sharp \cC(X)$. There are always finitely many extremal points, illustrated by the red dots in the panel on the right. The red dots define the $\sharp$complexity of $X$.\label{fig:FD}}
\end{center}
\vspace*{-.8cm}
\end{figure}

This leads us to view the format and degree $(F,D)$ as an effective measure of geometric complexity. However, a striking conceptual challenge is that these numbers are not uniquely assigned to definable sets. Consider any definable set $X\subseteq \bbR^m$. This set may for instance be the graph of a function, or the solution set of a system of equations and inequalities. The filtration property of $\Omega$ implies that there will be infinitely many pairs $(F,D)$ for which $X\in \Omega_{F,D}$. This situation may be visualized as in figure \ref{fig:FD}. In fact, this issue is no different than for Pfaffian complexity, where a Pfaffian function can be represented in many different ways. As a way of obtaining a uniquely defined complexity, one may consider to first minimize the format $F$ and then the degree $D$, or vice versa. However, in the present context, where $F$ and $D$ determine the complexity of computations applied to definable sets, the optimal choice of $(F,D)$ will depend in a case-by-case manner on the performed computation.

Instead of viewing the apparent non-uniqueness of format and degree as an obstacle, one should view this as a fundamental feature of the very notion of complexity: the complexity of an object is not purely inherent to the object itself, but also determined by the situation in which the object appears.\footnote{We are grateful to Gal Binyamini for stressing this point to us.} Therefore, it is natural to propose that the sharp complexity, or $\sharp$complexity for short, of a definable set $X$ should be defined as the collection of all formats and degrees to which $X$ belongs, i.e.~the set $ \{(F,D) \,|\, X\in \Omega_{F,D}\}$. This is an infinite set, but in fact only finitely many points are relevant. These are the `extremal' points $(F,D)$ with the property that
\begin{equation}
    X \notin \Omega_{F,D-1} \quad\text{and}\quad X \notin \Omega_{F-1,D} \,.
\end{equation}
The relevance of these points is that there is no trivial way to reduce $(F,D)$, and hence one of these points will lead to the optimal computational complexity when performing an operation on $X$. With this in mind, we define the $\sharp$complexity of $X$ as the finite set
\begin{equation}
    \sharp \cC(X) = \big\{(F,D) \,|\, X\in \Omega_{F,D}, \,X\notin \Omega_{F,D-1},\, X\notin \Omega_{F-1,D} \big\} \,.
\end{equation}

To illustrate the concept of $\sharp$complexity, let us compute it for a number of elementary examples. For concreteness, let us focus on the sharply o-minimal structure $\bbR_\text{alg}$, where the degrees of sets are determined by the degrees of underlying polynomials. First, consider a set $\{a\}$ consisting of a single point in $\bbR$. This is an algebraic set, namely the zero set of the polynomial $P(x)=x-a$. It follows from axiom (iv) of sharp o-minimality that $\{a\}\in \Omega_{1,1}$, and therefore $\sharp \cC(\{a\})=\{(1,1)\}$. More generally, consider a set of $N$ points $\{a_1,\ldots,a_N\}$ in $\bbR$. This set can be obtained as the zero set of a degree $N$ polynomial, i.e.~we have
\begin{equation}
    \{a_1,\ldots,a_N\} = \left\{ \prod_{j=1}^N(x-a_j)=0 \right\}\,.
\end{equation}
It thus follows that $\{a_1,\ldots,a_N\}\in\Omega_{1,N}$. In general, whether $(1,N)$ is part of the $\sharp$complexity of this set depends on the underlying sharply o-minimal structure. Since we are currently focusing on $\bbR_\text{alg}$, the definable sets can only be constructed by polynomials, so in this structure we have $(1,N) \in \sharp\cC(\{a_1,\ldots,a_N\})$. In other structures, e.g.~a Pfaffian structure, there may be parametrizations with lower complexity. Now let us increase the format $F$. Depending on the number of points $N$, there are alternative ways of obtaining this set by projecting a higher-dimensional set of lower degree. By B\'ezout's theorem, the intersection of $n$ degree $d$ hypersurfaces in $\bbR^n$ consists of at most $d^n$ isolated points. Projecting this down to the real line preserves the format and degree, so if $N\leq d^n$ we have $(n,nd) \in \sharp \cC(\{a_1,\ldots,a_N\})$. This example shows that even for the simplest sets, i.e.~finite collections of points, the $\sharp$complexity consists of several optimal pairs $(F,D)$.

As another example, consider a function $f$ definable in a sharply o-minimal structure. We then define the $\sharp$complexity of $f$ to be the $\sharp$complexity of the graph of $f$, i.e.~$\sharp\cC(f)=\sharp\cC(\Gamma_f)$. For instance, if $f$ is a degree $d$ polynomial in $n$ variables, then the graph $\Gamma_f$ is the zero set of the polynomial 
\begin{equation}
 P(x_1,\ldots,x_n,y)=y-f(x_1,\ldots,x_n).
\end{equation}
It follows that $(n+1,d)\in \sharp \cC(f)$. As a more non-trivial example, for Pfaffian functions the degree and format can be computed in terms of their Pfaffian complexity using the results of \cite{binyamini2022sharply}.

\subsection{Conjectures: Noetherian functions and periods} \label{sharpConjectures}

In reference \cite{BinNovICM}, it was pointed out that it is expected that all o-minimal structures which arise naturally in geometric settings should in fact be sharply o-minimal. These expectations were made precise by the following conjecture:
\[
\textbf{Conjecture. }\text{The structures }\bbR_\text{Pfaff},\,\bbR_\text{rNoether},\, \bbR_\text{Qf}\text{ are sharply o-minimal.} 
\]
To explain the meaning of this statement, we now review the definitions of the structures $\bbR_\text{rNoether}$ and $\bbR_\text{Qf}$ which we have not encountered before. First, let us introduce the concept of \textit{Noetherian functions}. The definition is almost identical to that of Pfaffian functions, the only difference being that the triangularity requirement of a Pfaffian chain is omitted. Explicitly, given an open box $U\subseteq \bbR^n$, i.e.~a product of open intervals, a Noetherian chain consists of functions $\zeta_1,\ldots,\zeta_r:U\to\bbR$ satisfying a system of differential equations of the form 
\begin{equation}
\frac{\partial \zeta_i}{\partial x_j} =  P_{ij} (x_1,\ldots,x_n,\zeta_1,\ldots,\zeta_r)  \quad\text{for} \quad i=1,\ldots,r, \quad j=1,\ldots, n \,.
\end{equation}
Analogous to Pfaffian functions, a Noetherian function is then defined by a polynomial in the variables $x_1,\ldots,x_n$ and the functions $\zeta_1,\ldots,\zeta_r$, and a Noetherian function is called restricted if $U$ is bounded. 

The structure generated by the Noetherian functions is denoted by $\bbR_\text{rNoether}$. It is known that the structure of restricted analytic functions is o-minimal, and since every Noetherian function is restricted analytic it follows that $\bbR_\text{rNoether}$ is o-minimal. In \cite{GabVor04}, it was shown that Noetherian functions exhibit various bounds on topological quantities. These bounds are analogous to the complexity bounds for Pfaffian functions, and this idea has led Binyamini and Novikov to conjecture that the structure $\bbR_\text{rNoether}$ is sharply o-minimal \cite{BinNovICM}. This conjecture is under active investigation. Since Noetherian functions are omnipresent, a consequence of this conjecture would be that the notion of $\sharp$-complexity extends far beyond the realm of Pfaffian complexity. We will illustrate in the next subsection, where we revisit the physical examples studied in section~\ref{sec:quantumsystems}.

It should be noted that one shortcoming of the class of restricted Noetherian functions is that their domains are bounded. This constraint is crucial, since general Noetherian functions are not necessarily tame (consider, for instance, the sin and cos function from section~\ref{sec:pfaffian_chains}). However, in geometric settings one naturally encounters tame functions on unbounded domains. It is expected that these functions also form a sharply o-minimal structure, but to make this precise one has to specify the class of functions under consideration. A candidate class is given by the so-called \textit{Q-functions}, whose exact definition we defer to Appendix \ref{app-Qf} due to their technical nature. Essentially, they arise as flat sections of connections on punctured polydiscs whose singularities are sufficiently regular. This class includes all period integrals of algebraic families. It is known that the structure generated by $Q$-functions, denoted by $\bbR_{Q\text{f}}$, is contained in the larger o-minimal structure $\bbR_\text{an,exp}$, which implies that they are tame. There are results which suggest that the functions in this structure obey certain complexity bounds \cite{BinNovYak}, which has led Binyamini and Novikov to conjecture that the structure $\bbR_{Q\text{f}}$ is sharply o-minimal.

\section{Sharp complexity in quantum systems} 
\label{sec:sharp_physics}

Now that we have reviewed the framework of sharp o-minimal structures and introduced the notion of $\sharp$complexity, we return to the physical theories studied earlier and reexamine how complexity manifests itself. Since the framework of sharp o-minimality is under active development, the following section will be more speculative in nature than our discussion of the more established Pfaffian complexity. 

\subsection{Lattice QFTs revisited}
In section~\ref{sec:quantumsystems}, we analyzed lattice quantum field theories, focusing on the extremely simple case of a one-point lattice and $\phi^4$- and $\phi^6$-interactions. We now turn to a far more general setting, and consider a scalar field theory on a finite lattice with a generic action. Our aim is to demonstrate, with the help of recent results found in \cite{Weinzierl:2022mmp}, that there exists a general systematic procedure for constructing a Noetherian chain containing the correlation functions of the theory. To precisely specify our setting, we fix a $D$-dimensional Euclidean lattice $\Lambda$ consisting of $N$ points, and consider a theory described by the action
\begin{equation}\label{eq:latticeaction}
    S=\sum_{x\in \Lambda}\left(-\sum_{\mu=0}^{D-1}\phi_x\phi_{x+  e_\mu}+D\phi_x^2+\sum_{j=2}^{J}\frac{\lambda_j}{j!}\phi_x^j\right)\;.
\end{equation}
Here the $x$ labels the lattice sites, $\phi_x$ is the value of the scalar field on the site $x$, and $e_\mu$ are basis vectors for the lattice $\Lambda$. The action includes the most general polynomial interaction term, and $J$ indicates the highest order interaction. Note that the nearest neighbour interactions of the form $\phi_x \phi_{x+e_\mu}$ encode the kinetic term of the scalar. The correlation functions of the theory now take the form of the following lattice path integral,\footnote{The domain of integration is $\cC^N$, where $\cC$ is a curve in the complex plane chosen such that the integrand vanishes at the boundary; this choice of contour enables us to integrate by parts.} 
\begin{equation}\label{Inu}
    I_{\nu_1\nu_2\cdots \nu_N}=\int\mathrm{d}^N \phi \left(\prod_{k=1}^{N}\phi_{x_k}^{\nu_k}\right)e^{-S} \,.
\end{equation}
Using techniques of twisted cohomology, it was recently argued in \cite{Weinzierl:2022mmp} that there exists a finite basis $I_1,\ldots,I_{n}$ of correlation functions, such that every correlation function can be written as a linear combination of this basis, with coefficients given by Laurent polynomials in the couplings $\lambda_1,\ldots,\lambda_J$. Moreover, the number of basis vectors $n$ is at most $(J-1)^N$. It was then shown that these basis integrals satisfy a system of differential equations of the form 
\begin{equation}\label{eq:Noether}
    \frac{\pd}{\pd \lambda_j}I_i=\sum_{k=1}^{N_F}A_{ik} I_k,
\end{equation}
for some square matrix $A$. The entries of $A$ are Laurent polynomials in the couplings. This essential fact about the structure of correlation functions in lattice QFTs can now be used to construct a Noetherian chain for which all correlation functions are Noetherian functions. Starting with the functions $f_1,\ldots,f_J$ defined by 
\begin{equation}
    f_j(\lambda_1,\ldots,\lambda_J)  = \frac{1}{\lambda_j} .
\end{equation}
The entries of the matrix $A$ are now polynomial in the couplings $\lambda_1,\ldots,\lambda_J$ and the functions $f_1,\ldots,f_J$. This means that the system of differential equations in \eqref{eq:Noether} forms a Noetherian chain. In fact, that $I_1,\ldots,I_n$ form a basis implies that all correlation functions in the theory are Noetherian with respect to this chain. Note that, if the matrix $A$ were triangular, this chain would in fact have been Pfaffian. In general, $A$ is rarely triangular, meaning that the Pfaffian framework cannot be implemented. 

If we now restrict the space of couplings $\lambda_1,\ldots,\lambda_J$ by imposing bounds on each $\lambda_j$, then the correlation functions $I_{\nu_1\cdots \nu_N}(\lambda_1,\ldots,\lambda_j)$ are restricted Noetherian functions and therefore definable in the o-minimal structure $\bbR_{\text{rNoether}}$. As explained in the previous section, this structure is expected to be sharply o-minimal. A consequence of this conjecture is that our notion of $\sharp$-complexity can be implemented for any lattice QFT with an action described by an action of the form in equation \eqref{eq:latticeaction}. In particular, all correlation functions carry a format $F$ and degree $D$, whose dependence on the parameters of the theory such as the dimension and size of the lattice, and the highest interaction $J$ could be analyzed. 

Let us emphasize that the preceding discussion is only valid upon restricting the space of couplings. This means that, in the restricted Noetherian setting, one loses the ability to probe potentially interesting infinite coupling limits. In contrast, the Pfaffian framework is globally valid. It would therefore be interesting to analyze whether the correlation functions reside inside a different structure which includes infinite domains while retaining the sharp complexity bounds. A natural candidate would be the structure of $Q$-functions, which is conjectured to be sharply o-minimal as explained in the previous section. We leave a more detailed analysis of lattice QFTs and sharp o-minimality for future work.

\subsection{Quantum mechanics revisited}
We now return to the quantum mechanical theories discussed in section~\ref{QuantumMPfaff}. In particular, we focus on the complexity of the wavefunctions arising as solutions to the Schr\"odinger equation. For a general polynomial potential, we were only able to show that the ground state is a Pfaffian function. As we will argue below, the structure $\bbR_{\rm rNoether}$ opens the door to considering the complexity of any wavefunction, as long as one restricts to a bounded region of space. Since we are interested in bound states which are localized near the minima of the potential, this provides an adequate description of the full wavefunction.  

More precisely, let $V(x)$ be an even-degree polynomial potential and consider the Schrödinger equation 
\begin{equation}
    -\frac{1}{2m} \psi''(x) + \big(V(x)-E\big)\psi(x) =0\, .
\end{equation}
It can directly be transformed to a Noetherian chain as follows
\begin{align}
    \zeta_1(x)& = \psi(x)  &&  \frac{\pd\zeta_1}{\pd x}= \zeta_2 \,, \\
    \zeta_2(x)&=\psi'(x)   && \frac{\pd\zeta_2}{\pd x}= 2m(V-E)\zeta_1\,. \nonumber
\end{align}
This shows that any wavefunction $\psi$ solving the Schr\"odinger equation, restricted to a bounded interval $(a,b)\subseteq\bbR$, is definable in $\bbR_{\rm rNoether}$. An analysis of the $\sharp$complexity of wavefunctions would require a better understanding of the sharp o-minimality of the structure $\bbR_{\rm rNoether}$, which is currently not known in the literature.

\subsection{Scattering amplitudes and effective theories}
\label{SWrevisited}
Many physical quantities are geometric in origin, and the idea that such geometric functions should belong to a sharply o-minimal structure as conjectured in \cite{BinNovICM} suggests that it is possible to assign a $\sharp$complexity to them. Below we comment on a few notable instances where these functions arise in physics, which provide interesting avenues to explore in the future.

\subsubsection*{Scattering amplitudes}

It was recently shown in  \cite{Douglas:2022ynw} that perturbative QFT scattering amplitudes at finite loop order are functions definable in $\bbR_{\rm an,exp}$, when considering their dependence on external momenta, the  
masses, and the coupling constant. This result uses the fact that 
these finite-loop amplitudes are related to period integrals \cite{Weinzierl:2022eaz}.  
However, it is not sufficient to assign a complexity, since $\mathbb{R}_{\rm an,exp}$ is not $\sharp$o-minimal. Recalling the conjectures of section~\ref{sharpConjectures} we know that the period integrals are also belonging to a more restricted class of functions which is contained in the structure of $Q$-functions, $\mathbb{R}_{Q\rm f}$. Thus, similar to the 0-dimensional case, there is a natural complexity associated to perturbative amplitudes given by the format and degree of the relevant Q-functions.

\subsubsection*{Seiberg-Witten theory revisited}
To go beyond perturbative QFT, one typically requires more symmetries which render the theory tractable. As we have seen before, one such example is Seiberg-Witten theory. In our discussion in section~\ref{sec:SW}, we focused on the case where the gauge group is given by $\text{SU}(2)$, but the theory may also be formulated for higher-rank gauge groups. For instance, in the case of $\text{SU}(N)$ the Seiberg-Witten curve is replaced by a hyperelliptic curve of genus $g=N-1$. This curve may be described by the equation 
\begin{equation} \label{SW-curve-N}
    y^2=\left(x^N-\sum_{i=2}^Nc_i x^{N-i}\right)^2-\Lambda^2N\,,
\end{equation}
where $c_i$ are the Casimir operators of the gauge group. The coupling functions of the theory are now described by period integrals of this curve, which are contained in the class of Q-functions. As a result, the coupling functions of Seiberg-Witten theory are conjecturally definable in the sharply o-minimal structure $\bbR_\text{Qf}$, which means that they have an associated $\sharp$complexity. We expect that the $\sharp$complexity of the coupling functions of the theory then increases with the rank $N$. 

While we leave a precise analysis of the scaling of $(F,D)$ with $N$ for future research, let us briefly highlight some  facts
about periods of Riemann surfaces that are crucial in this context. 
Concretely, we want to argue that the format $F$ is expected to grow with $N$.\footnote{This should be contrasted with the fact that, for example, for correlation functions studied in section \ref{QFTsonPoint}, we see no such growth with the number of operator insertions due to the algebraic relations among correlators.} 
Clearly the number of periods increases with the genus of the Riemann surface, since there are more cycles to integrate over. 
Using the recent theorem \cite{BTnew} one realizes that at higher $g$, and hence higher $N$, more and more algebraically independent periods arise from the Seiberg-Witten curve.\footnote{We would like to thank Benjamin Bakker for discussions on this argument.} Hence, one will not be able to find a universal basis of functions encoding the 
corrections for all Seiberg-Witten curves but rather one will see a growth of $F$ with the rank $N$ of the SU($N$).

\subsubsection*{Effective field theories from type II string theory}

It has been shown that Seiberg-Witten theories can be obtained from string theory on certain non-compact manifolds constructed from the Seiberg-Witten curve \cite{Klemm:1996bj}. More generally, the geometrization of coupling functions of an effective theory arising from string theory is present for many string compactifications. For example, the period functions belonging to the structure $\bbR_\text{Qf}$ feature prominently in effective theories arising from Calabi-Yau compactifications of type IIB string theory and F-theory. This suggests that it may be possible to assign a $\sharp$complexity to these effective theories, which populate a large part of the string landscape. We will comment on this further in the discussion, and address this idea in an upcoming work \cite{GrimmVliet}.

\section{Conclusions and discussion}

In this work, we were driven by the foundational premise that physical systems inherently possess a finite degree of complexity or information content. To systematically encapsulate this idea, we leaned on tame geometry, which is built by using o-minimal structures, as a mathematically rigorous way to assign complexity. The first concrete realization of this idea is realized when using Pfaffian functions and Pfaffian sets, both of which exhibit a well-defined notion of complexity, that we termed Pfaffian complexity. We have shown that several physical systems can be described using these functions and hence are amenable to this approach. Even without referring to a notion of complexity, the use of Pfaffian functions turned out to be a hands-on way to prove the tameness property of physical systems that goes beyond the analysis carried out for the examples in \cite{Douglas:2022ynw,Douglas:2023fcg}. Interestingly, any slight generlizations of our physical examples eventually pointed to the necessity of also generalizing the 
mathematical description beyond Pfaffian structures. This led us to consider sharp o-minimal structures, which were recently introduced \cite{BinNovICM,binyamini2022sharply} as a mathematically well-motivated overarching framework generalizing the Pfaffian setting. 

The definition of sharp o-minimality implements a well-defined notion of complexity by associating sets of two integers, the format $F$ and the degree $D$, to each set or function defined in such a structure, which we termed $\sharp$complexity. This notion has many remarkable properties. In particular, the definition guarantees that whenever a set of degree $D$ is intersected with the real line, the resulting number of connected components depends polynomially on $D$. This implies that at least for fixed format $F$, the computational effort to check a logical statement is polynomial in $D$. This property is compatible with all logical operations and makes the finiteness statement of o-minimality a quantitative attribute. It can indeed be understood as introducing a well-defined notion of the amount of information needed to logically define a function or a set. 
The fact that there are generally multiple minimal pairs $(F,D)$ associated to a set or function should be regarded as a feature instead of a bug; it indicates that the set admits different equivalent representations, in a similar spirit as dualities.

In order to illustrate the notion of Pfaffian complexity and $\sharp$complexity in physical systems we 
focused on three qualitatively different classes of examples.  
We first discussed $\phi^n$ theories on a finite number of points. For the simplest settings, namely $\phi^4$ and $\phi^6$ theories on a point, we showed that both correlation functions and partition functions can be written as Pfaffian functions, thereby enabling the assignment of a Pfaffian complexity to them. Interestingly, we observed that the complexity grows linearly with the number of field insertions. 
We also observed an increment in complexity when transitioning from $\phi^4$ to $\phi^6$ theory, leading us to infer a complexity growth for $\phi^n$ as $n$ increases. However, we have seen that beyond the simple $\phi^4,\phi^6$ a more involved sharp o-minimal structure seems to be necessary. Using the notion of $\sharp$complexity $(F,D)$, we expect that for $\phi^n$ theories on any number of points, the degree $D$ increases with the number of insertions in a correlator, while the format increases with $n$. At least when restricting the domains of the parameters, we were able to identify a candidate sharply o-minimal structure, $\bbR_{\rm rNoether}$, in which one can try to make these statements precise. 

Another striking observation, which strengthens the statements of \cite{Douglas:2022ynw,Douglas:2023fcg}, was the fact that Pfaffian structures enable us to treat functions which are not analytic but rather only admit a transseries expansion and are experiencing resurgence phenomena. Given the fact that such functions are ubiquitous in quantum field theories \cite{Dorigoni:2014hea}, we believe that the broadened scope presented in this work will be instrumental in the future.

Also in settings of quantum mechanics, Pfaffian functions arise naturally. Specifically, we showed that in a one-dimensional setting, the wave functions are, at least for simple potentials, Pfaffian functions and admit a Pfaffian complexity related to the energy level. While for general potentials the ground state remains Pfaffian, one has to generalize to a sharply o-minimal setting when addressing higher-level wave functions. Clearly, our analysis should be only seen as an initial step of implementing a new notion of complexity in quantum mechanics. These ideas need to be compared 
to existing notions of complexity in this context. 
We have made some first remarks on the relationship between Pfaffian complexity and quantum computational complexity, but hope 
that a more comprehensive study 
will be carried out in the future. In addition, it would be worthwhile to explore connections with sharp complexity and holography, along the lines of the notion of holographic complexity \cite{Susskind:2014rva,Brown:2015bva}. An intriguing facet of the Pfaffian complexity and $\sharp$complexity is 
that certain information about a function, such as the coefficients in a polynomial, enters the derived complexity when representing the function in a specific way. Here one might follow a lesson from quantum computational complexity
and assert that also the coefficients in a polynomial should inherently possess a complexity and be `quantized' in an appropriate way. To implement this explicitly remains an interesting open problem.

Finally, we analyzed couplings functions in a special class of quantum field theories, namely Seiberg-Witten theories, and commented on the complexity of general scattering amplitudes. The Seiberg-Witten example was meant to give a first indication of how Pfaffian complexity or $\sharp$complexity can be associated to an effective QFT. We primarily examined the gauge-coupling functions in these theories and used the fact that, due to supersymmetry constraints, they can be non-perturbatively derived from an auxiliary geometry. 
For an $\text{SU}(2)$ gauge group without hypermultiplets we found that the Pfaffian setting suffices to give a partial description, while again the use of sharply o-minimal structures becomes immediately relevant when trying to generalize further. Also in this example, we found that the complexity highly depends on the representation of the physical quantity, e.g.~via a choice of coordinates. The challenge hereby is that a focus on a single physical quantity within an effective theory might lead to a complexity that is artificially low, since it does not take into account the intricacies of the complexity minimization problem relevant when considering the full theory. 

Note that our explicit analysis was focused largely on the case where the gauge group is $\text{SU}(2)$. However, we presented first arguments for how the $\sharp$complexity of the gauge coupling functions of SU$(N)$ Seiberg-Witten theory will grow with $N$. Deriving this complexity-growth explicitly and examining the assignments to other coupling functions in this setting is left as an important open problem for the further research. More generally, it will be crucial to investigate $\sharp$complexity of effective field theories and we hope to return to this in the future.

\noindent
\textbf{On the information-theoretic perspective.}
This work was motivated by setting the notion of `finiteness of information' as a core principle that needs to be obeyed in a physical theory. Let us stress that our current understanding is insufficient to present a full story on this point. The preceding discussion on complexity of coefficients in a polynomial already highlights the difficulty of this problem. Indeed, even the choice of a number, say the real exponent $\alpha$ of the functions $x^\alpha$, can contain an infinite amount of information if $\alpha$ is irrational and not determined by a `simple' equation.\footnote{For example, $\alpha$ could be an algebraic number.} A general sharply o-minimal structure does not record all the information in the coefficients of the defining equations of the sets, even though it is also not independent of it either. Our current approach can be labelled as setting the focus on the functional complexity, rather than the complexity of numbers.
To give an analogy to recent studies in mathematics, we note that vast progress on Ax-Schanuel-type theorems, studying transcendentality properties of functions, has been made using o-minimality, see e.g.~\cite{AxSchanuelHodge}. In contrast, the Schanuel conjecture and the period conjecture \cite{Kontsevich2001}, suggesting transcendentality properties of numbers, is still wide open.\footnote{It is worth noting that \cite{Kontsevich2001} features a brief discussion on complexity of numbers.} We see the search for a refined framework within mathematics and physics as the most central challenge to realize the general principle of `finiteness of information'.

\noindent
\textbf{On the coupling to gravity.} In this work we have considered quantum systems that are not coupled to gravity. It is an interesting question in how far tameness, with the sharpened notion discussed in this work, is related to a coupling to gravity. It was suggested in \cite{Grimm:2021vpn}, and further explored in \cite{Grimm:2022sbl,Douglas:2023fcg}, that o-minimality is a necessary condition for an effective theory to admit a UV completion with gravity. It is not expected, however, that it is also a sufficient condition. To see this, consider, for example, $\text{SU}(N)$ Seiberg-Witten theories. For very large $N$ we do not expect that these can arise from a string theory compactification with a finite four-dimensional Planck mass, and one might be inclined to conclude that such theories can never be coupled to quantum gravity. However, the o-minimality property of $\text{SU}(N)$ Seiberg-Witten theories, e.g.~of their coupling functions, does not seem depend on $N$ and, in fact, we expect that all these theories are tame. Remarkably, within the context of sharp o-minimality, we now have a quantity that does depend on $N$, namely the $\sharp$complexity discussed in section~\ref{SWrevisited}. Hence, we expect that theories compatible with quantum gravity will have to obey a complexity bound. In an upcoming work \cite{GrimmVliet} we 
will make this more precise.

\subsubsection*{Acknowledgements}

We profited immensely from discussions with Lou van den Dries and Gal Binyamini and would like to thank them for sharing their insights and understanding with us. Furthermore, would like to thank Benjamin Bakker, Michael Douglas, Gerard~'t~Hooft, Damian van de Heisteeg, Ro Jefferson, and Cumrun Vafa for useful discussions and comments. This research is supported, in part, by the Dutch Research Council (NWO) via a Vici grant.

\appendix

\section{Cell decomposition in tame geometry} \label{app-cells}

In this appendix we briefly describe the cell decomposition theorem, mainly following \cite{VdDries}. The following discussion holds for any o-minimal structure $\cS$, and we will keep this implicit in the notation.

Let us start by introducing a more basic version of the cell decomposition theorem, called the monotonicity theorem. Consider a real-valued definable function $f$ defined on a (possibly infinite) open interval $(a,b)$. The monotonicity theorem states that the interval $(a,b)$ can be subdivided into finitely many pieces on which $f$ is either monotonic and continuous or constant. More precisely, there exist points 
\begin{equation}\label{split-int}
    a=:a_0 < a_1 < ...< a_{m-1} < a_m:=b\ ,
\end{equation} 
such that $f$ is either constant, or strictly monotonic and continuous on the open intervals $(a_k,a_{k+1})\subseteq (a,b)$. Consequently, $f$ can change from decreasing to increasing only finitely many times, thereby ruling out infinite oscillations. Moreover, $f$ can only have finitely many discontinuities. A stronger version of this theorem ensures that $f$ is $C^p$-differentiable on (possibly smaller) open intervals in a finite decomposition. The subintervals $(a_k,a_{k+1})$, along with the individual points $a_k$, may be thought of as \textit{cells} in the interval $(a,b)$ on which the behavior of $f$ is particularly simple.

Next, let us introduce the cell decomposition of $\bbR^n$. This is done inductively, with the cell decomposition on the real line $\bbR$ as described above taken as a starting point. In higher dimensions, cells are defined to be graphs or bands which are delimited by definable functions in one dimension lower. The details are as follows. For $n>0$, we write $\bbR^n = \bbR^{n-1} \times \bbR$. Suppose that $\bbR^{n-1}$ is decomposed into cells $\{ \cC_\alpha\}$. One introduces for each cell $\cC_\alpha$ an integer $m_\alpha>0$ and a set of definable continuous functions $f^{(\alpha)}_k: \cC_\alpha \rightarrow \bbR$ for $0< k < m_\alpha$ 
such that on the entire domain $\cC_\alpha$, we have
\beq
   -\infty =: f^{(\alpha)}_0  < f^{(\alpha)}_1 < \ldots < f^{(\alpha)}_{m_\alpha -1} < f^{(\alpha)}_{m_{\alpha}} := \infty\ ,  
\eeq
From this set of functions, the cells in $\bbR^n$ are given by: 
\begin{itemize}
    \item[(i)] graphs of functions:
   $\{(x,f^{(\alpha)}_k(x))\subset \bbR^n:x\in \cC_\alpha \}$ for each 
$\cC_\alpha$;  
\item[(ii)] bands between functions: 
$\{(x,y)\subset \bbR^n: x\in \cC_\alpha, y\in (f^{(\alpha)}_k(x),f^{(\alpha)}_{k+1}(x))\}$.
\end{itemize}

\begin{figure}[h!]
\begin{center}
\vspace*{.5cm}
 \includegraphics[width=0.55\textwidth]{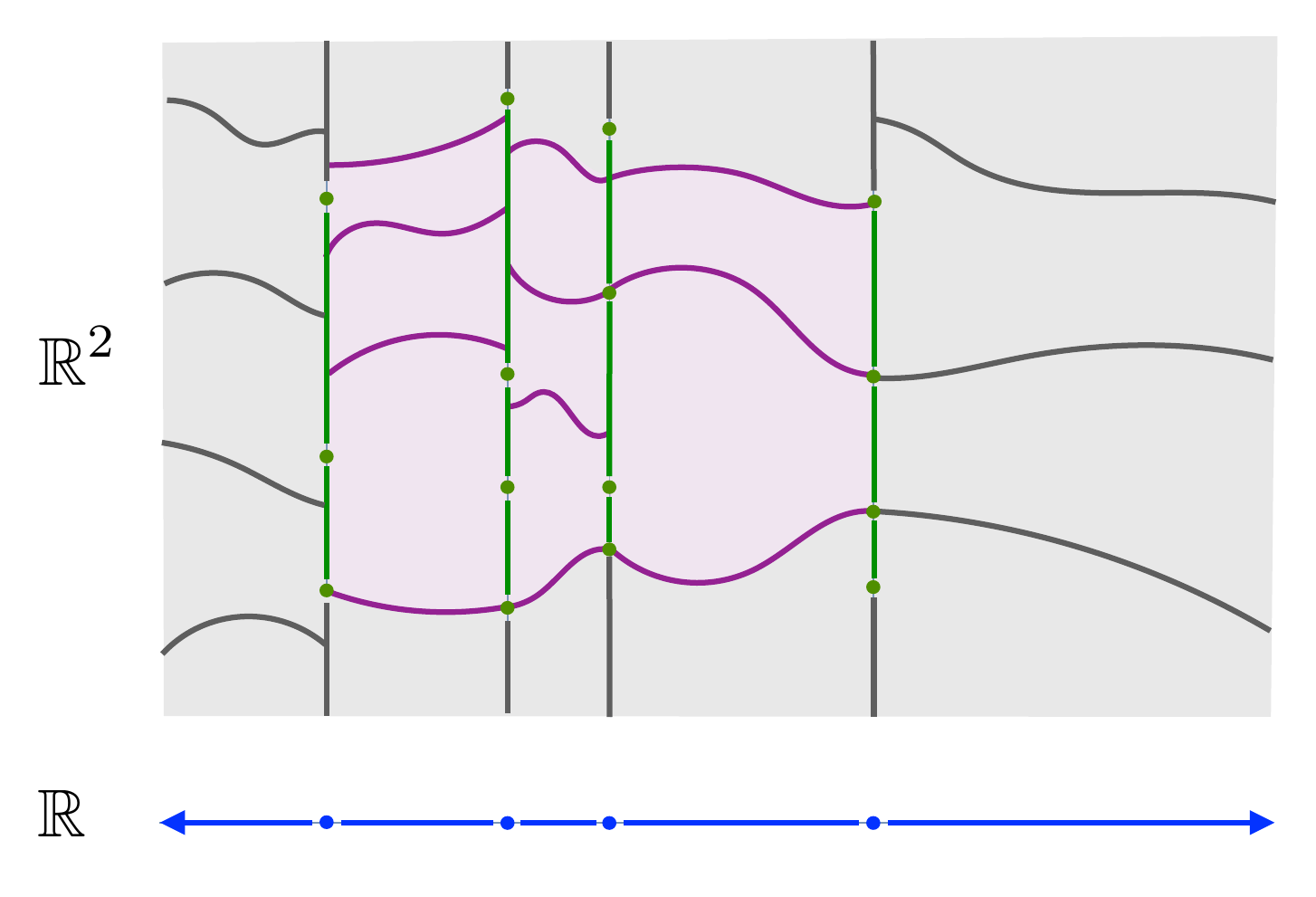} 
 \vspace*{.1cm}
\caption{An example of cell decomposition of $\bbR^2$ (above), constructed by a cell decomposition of $\bbR$ (below). The cells of $\bbR^2$ are graphs and bands of continuous definable functions on the $\bbR$-cells. The new cells over open intervals in $\bbR$ are shown in purple, while the new cells over points in $\bbR$ are shown in green. 
The grey cells should be thought of as extending to infinity.  \label{def-decR2}}
\end{center}
\vspace*{-.8cm}
\end{figure}

The result of this construction is a decomposition of $\bbR^n$ into a finite disjoint collection of definable subsets. The fact that they are constructed in terms of continuous definable functions ensures that cells have a simple geometry. As with the monotonicity theorem, there also exists a refined notion of cell for which the functions are assumed to be $C^p$ differentiable. A subtle but important technical fact is that, for every $n$, there is a preferred projection $\bbR^n\to\bbR^{n-1}$ along which the boundaries of the cells are straight. Because of this, the cell decomposition is often called `cylindrical'. Figure \ref{def-decR2} shows an example of part of a cell decomposition of $\bbR^2$.

The \textit{cell decomposition theorem} now states that definable cells are sufficient to describe any definable set. The precise statement is as follows:
\begin{itemize}
    \item[(i)] For any finite collection of definable sets $A_1,...,A_k \in \bbR^n$ there exists a cell decomposition such that each $A_i$ is a finite union of cells;
    \item[(ii)] For each definable function $f:A\rightarrow \bbR$, with $A\subset \bbR^n$, there is a cell decomposition of $\bbR^n$, partitioning $A$ as in part (i), such that the restriction of $f$ to any cell is continuous.
\end{itemize}
The intuition for point (i) is that the geometry of any definable set may be described in terms of finitely many simple pieces. In technical applications of tame geometry, this means that geometric problems of a definable set can be reduced to finitely many cells. Applications of this form rely on the fact that there exist explicit geometric algorithms which find the cell decomposition of a given definable set. Point (ii) may be seen as a higher-dimensional analogue of the monotonicity theorem.

The cell decomposition theorem admits a refinement upon specializing to sharply o-minimal structures \cite{binyamini2022sharply}.\footnote{To be precise, it was shown that any sharply o-minimal structure admits a refined FD-filtration for which the sharp cell decomposition theorem holds.} The crucial difference is that the \textit{sharp} cell decomposition is more quantitative in nature, in the sense that the number of cells and the formats and degrees of the cells are controlled in a polynomial manner by the formats and degrees of the sets $A_1,\ldots,A_k$.

\section{$Q$-functions} \label{app-Qf}
Here we briefly elaborate on the class of $Q$-functions, which generate the conjecturally sharply o-minimal structure $\bbR_{Q\text{f}}$. We follow the exposition in \cite{BinNovICM}. The idea behind this class of functions is that they are defined on unbounded domains, while still having sufficiently controlled singularities at infinity. They arise naturally in geometry, for instance in Hodge theory. The precise definition is as follows. 

First, let $D\subseteq \bbC^n$ be a polydisk and let $\Sigma\subseteq \bbC^n$ be a union of coordinate hyperplanes. On the vector bundle $D\times \bbC^l$, we consider the connection 
\begin{equation}
    \nabla v = dv - A\cdot v,
\end{equation}
where $A$ is a matrix of one-forms with the following properties:
\begin{enumerate}
    \item[(i)] the one-forms in $A$ are holomorphic on $\overline{D}\setminus \Sigma$, where $\overline{D}$ is the closure of $D$;
    \item[(ii)] the entries of $A$ are algebraic over the field of algebraic numbers $\overline{\mathbb{Q}}$;
    \item[(iii)] the connection $\nabla$ has regular singularities along $\Sigma$;
    \item[(iv)] the connection $\nabla$ has quasi-unipotent monodromy.
\end{enumerate}
To eliminate the effects of monodromy, we remove a branch cut from $D$. In this way, we obtain a simply connected domain $D^\circ$. The $Q$-functions is now defined as the component functions of flat connections of $\nabla$, i.e.~solutions to $\nabla v=0$ on $D^\circ$. By construction, $Q$-functions are holomorphic, and in fact the constraints on $A$ guarantee that these functions are definable in the o-minimal structure $\bbR_\text{an,exp}$ and hence tame. It was shown that these functions obey certain complexity bounds similar to those in sharp o-minimality \cite{BinNovYak}, leading to the conjecture that the structure $\bbR_{Q\text{f}}$ generated by them is sharply o-minimal \cite{BinNovICM}.

%It should be noted that one shortcoming of the class of restricted Noetherian functions is that their domains are bounded. This constraint is crucial, since general Noetherian functions are not necessarily tame (consider, for instance, the sin and cos function from section~\ref{sec:pfaffian_chains}). However, in geometric settings one naturally encounters tame functions on unbounded domains. It is expected that these functions also form a sharply o-minimal structure, but to make this precise one has to specify the class of functions under consideration. A candidate class is given by the so-called \textit{Q-functions}, whose exact definition we defer to Appendix \ref{app-Qf} due to their technical nature. Essentially, they arise as flat sections of connections on punctured polydiscs whose singularities are sufficiently regular. This class includes all period integrals of algebraic families. It is known that the structure generated by $Q$-functions, denoted by $\bbR_{Q\text{f}}$, is contained in the larger o-minimal structure $\bbR_\text{an,exp}$, which implies that they are tame. There are results which suggest that the functions in this structure obey certain complexity bounds \cite{BinNovYak}, which has led Binyamini and Novikov to conjecture that the structure $\bbR_{Q\text{f}}$ is sharply o-minimal. 

\bibliography{literature}
\bibliographystyle{utphys}

\end{document}